\documentclass[12pt,preprint,prd,nofootinbib]{revtex4-1}

\usepackage[pdftex]{graphicx}
\usepackage{latexsym,amsmath,amssymb,lmodern,float,url}
\usepackage{natbib}
\usepackage[pdftex,bookmarks,linktocpage,pdfpagelabels,plainpages=false,hyperfigures,linkcolor=blue,citecolor=blue]{hyperref} 
\usepackage{xspace}
\hypersetup{colorlinks=true}

\newcommand\be{\begin{equation}}
\newcommand\ee{\end{equation}}
\newcommand\bea{\begin{eqnarray}}
\newcommand\eea{\end{eqnarray}}
\newcommand{\appropto}{\mathrel{\vcenter{
  \offinterlineskip\halign{\hfil$##$\cr
    \propto\cr\noalign{\kern2pt}\sim\cr\noalign{\kern-2pt}}}}}
\newcommand{\cns}{CE$\nu$NS\xspace}
\newcommand{\tmax}{\ensuremath{E_R^{\rm max}}\xspace}
\newcommand{\tmin}{\ensuremath{E_R^{\rm min}}\xspace}

\begin{document}
\preprint{MI-TH-1637}
\preprint{CETUP2016-008}

\title{Probing light mediators at ultra-low threshold energies with coherent elastic neutrino-nucleus scattering}
\author{James~B.~Dent$^{\bf a, b}$}
\author{Bhaskar~Dutta$^{\bf c}$}
\author{Shu~Liao$^{\bf c}$}
\author{Jayden~L.~Newstead$^{\bf d}$}
\author{Louis~E.~Strigari$^{\bf c}$}
\author{Joel~W.~Walker$^{\bf b, e}$}

\affiliation{$^{\bf a}$ Department of Physics, University of Louisiana at Lafayette, Lafayette, LA 70504, USA}
\affiliation{$^{\bf b}$ Kavli Institute for Theoretical Physics, University of California, Santa Barbara, CA 93106-4030, USA}
\affiliation{$^{\bf c}$ Mitchell Institute for Fundamental Physics and Astronomy,
   Department of Physics and Astronomy, Texas A\&M University, College Station, TX 77845, USA}
\affiliation{$^{\bf d}$ Department of Physics, Arizona State University, Tempe, AZ 85287, USA}
\affiliation{$^{\bf e}$ Department of Physics, Sam Houston State University, Huntsville, TX 77341, USA}

\begin{abstract}
Light neutral mediators, with mass $\lesssim 1$ GeV, are common features of extensions to the Standard Model (SM).
Current astrophysical and terrestrial experiments have constrained the model parameter space, and planned experiments 
around the world promise continued improvement in sensitivity. In this paper we study the prospects for probing light neutral 
mediators using terrestrial stopped pion and reactor sources in combination with ultra-low threshold nuclear and electron recoil detectors.  
We show that the coherent neutrino-nucleus and neutrino-electron scattering channels provide complementary sensitivity to 
light mediators. With low threshold detectors, we show that most stringent bounds on models arise from the nuclear 
scattering process, improving upon previous bounds from electron scattering of solar neutrinos by nearly an order of magnitude 
for mediator masses $\gtrsim 0.1$ GeV.

\end{abstract}

\maketitle

\section{Introduction}
Low mass particles from hidden sectors occur in many extensions of the Standard Model (SM).
Examples include grand unified theories (GUTs), models that explain baryogenesis, or dark sector
models that include a portal which provides a connection to SM particles.
Many experiments around the world are now being developed to study light mediators and hidden sectors
(see for example the review~\cite{Essig:2013lka}).

Experiments that are designed to detect coherent elastic neutrino-nucleus scattering (\cns) can provide
a particularly important probe of light mediator models.  Because the
\cns process is well predicted in the SM~\cite{Freedman:1973yd}, a measured deviation from it
can provide a test of Beyond Standard Model (BSM) physics. These models include
non-standard neutrino interactions (NSI) with new weakly coupled particles, electromagnetic properties
of neutrinos, and oscillations into sterile neutrinos~\cite{Krauss:1991ba,Barranco:2005yy,Barranco:2007tz,
Formaggio:2011jt,deNiverville:2015mwa,Kosmas:2015sqa,Kosmas:2015vsa,Dutta:2015vwa,Dutta:2015nlo,Lindner:2016wff}.
\cns can also shed light onto astrophysical processes occurring in neutron stars and core-collapse
supernovae~\cite{Wilson:1974zz,Prakash:2000jr,Horowitz:2003cz,Burrows:2004vq,Janka:2006fh}.

Although experiments to measure \cns~\cite{Drukier:1983gj,Cabrera:1984rr} have been long proposed,
its detection has remained elusive. However, recent experimental progress on several fronts may make
detection of \cns a reality in the near future. The COHERENT experiment will use a stopped pion beam
at the Spallation Neutrino Source (SNS) to detect \cns with Argon, Germanium, and CsI detectors~\cite{Akimov:2015nza},
and will improve upon current NSI limits. The CONNIE experiment is currently using low-threshold CCD detectors near a GW reactor source with near term prospects for measuring the CE$\nu$NS process~\cite{Aguilar-Arevalo:2016khx}, and the TEXONO experiment is developing detectors to measure \cns also using a
GW reactor neutrino source.  These experiments will be sensitive to BSM physics such as NSI and the neutrino magnetic moment~\cite{BinLi:2016vyy}.
More recently, the MINER experiment at Texas A\&M University has been developed to measure to measure \cns
using a MW reactor neutrino source, with sensitivity to sterile neutrinos, NSI, and the neutrino magnetic
moment~\cite{Dutta:2015nlo,Dutta:2015vwa}.

In addition to these terrestrial efforts, \cns will also be an important aspect of direct dark matter searches.
Direct detection experiments search for the interaction of dark matter (DM) with SM particles, typically via the nuclear recoil
induced by elastic DM-SM scattering.
As direct detection experiments continue to increase in sensitivity, the solar, atmospheric, and diffuse
neutrino background from \cns will become a pressing issue. For canonical parameterizations of the DM-nucleus
cross section, the presence of the \cns background will result in a situation where increasing the experimental exposure typically does not improve the sensitivity to lower cross-sections,
an obstacle known as the neutrino floor \cite{Billard:2013qya,Ruppin:2014bra}. However, for more general parameterizations of DM interactions,
the neutrino floor can be mitigated in a large fraction of the parameter space~\cite{Dent:2016iht,Dent:2016wor}.

Several papers have studied how \cns can be used to probe light mediators. Ref.~\cite{Harnik:2012ni,Cerdeno:2016sfi}
use solar neutrinos at direct detection experiments, studying dark photon models and generic light mediator interactions of scalar, pseudoscalar, vector and axial-vector types. Ref.~\cite{deNiverville:2015mwa} analyze the sensitivity of several \cns experiments to light dark matter interacting with the SM through a light vector mediator coupled to the electromagnetic current.

In this paper we study the prospects for identifying NSI with light neutral mediators using
a combination of \cns and neutrino-electron scattering experiments. New mediator particles whose
masses are small compared to the typical momentum exchanged in the scattering process can significantly
alter the scattering rate, with pronounced effects at low recoil kinetic energies ($\lesssim$ keV),
and are therefore an intriguing experimental target.
We focus specifically on projections for the MINER experiment and the first phase of the COHERENT project, using a simplified model approach to
project constraints on hidden sector models which include a new light vector boson that couples neutrinos to the
lepton and quark sectors of the SM.
We show that an experiment with the sensitivity of MINER can produce world-leading constraints on a $U(1)_{\mathrm{B-L}}$ extension of the SM. 
Interestingly, the most stringent bounds arise from the nuclear scattering process rather than electron scattering as in solar neutrino probes.
This is because the largest solar fluxes correspond to relatively low energy regions of the neutrino spectrum, 
with correspondingly soft nuclear recoils below current detection thresholds, whereas reactors combine moderate
neutrino energies with a large local flux.
The hope for experimentally leveraging advantages of the \cns process relies upon a continued
push towards lower threshold detector technologies, such as those actively being developed for direct DM searches.
This program is especially central to the study of models with light mediators, given the described rate
enhancement in the ultra-low recoil energy regime. 

The outline of this paper is as follows.
Sec.~\ref{sct:cnssm} briefly reviews some details of \cns in the SM, and
Sec.~\ref{sct:extensions} summarizes a generic simplified model extension of the SM that includes NSI via light mediators.
Sec.~\ref{sct:detectors} describes the application of very low threshold detector technology to reactor searches. 
Sec.~\ref{sct:smconflimits} briefly lays out the process employed for limit setting, and
Sec.~\ref{sct:limits} presents our main numerical sensitivity projections, including
results for a specific $U(1)_{\mathrm{B-L}}$ BSM model at MINER.
Sec.~\ref{sct:scaling} examines the relative scaling of these limits in the extreme light mediator limit.
In Sec.~\ref{sct:conclusions}, we summarize and conclude.

\section{\cns in the Standard Model\label{sct:cnssm}}

Coherent elastic neutrino-nucleus scattering has garnered great attention as an important process for astrophysics,
dark matter physics, and as a test of the Standard Model itself in a new, low energy regime.  The scattering proceeds
due to a neutrino of incident energy $E_{\nu}$ impinging on a nuclear target, to which is imparted a kinetic recoil
energy $E_R$.  Coherent scattering arises as a pure quantum effect for incident neutrinos with small enough energy
such that they are unable to probe the interior nucleon structure of the nucleus.
Specifically, for momentum transfers $|\vec{q}| \lesssim R^{-1}$,
with $R$ the typical nuclear size, coherent scattering will generate an enhancement in the cross-section.

The differential cross-section for a neutrino scattering off of
a target of mass $m$ (such as an electron or quark) is
\be
\frac{d\sigma}{dE_R} = \frac{G_F^2 m}{2\pi}\big((g_v+g_a)^2 +
(g_v-g_a)^2\left(1-\frac{E_R}{E_{\nu}}\right)^2 +
(g_a^2-g_v^2)\frac{mE_R}{E_\nu^2}\big)
\, .
\label{eq:dcs}
\ee
Where $G_F$ is the Fermi constant. The vector and axial-vector couplings
$(g_v,g_a) \equiv (g_L+g_R,g_L-g_R) = (T_3-2Q_{\rm em}{\rm{sin}}^2\theta_W,T_3)$
are those of the scattering target
with respect to the neutral current amplitude $(T_3-Q_{\rm em}\,{\rm{sin}}^2\theta_W)$ for
coupling to a $Z$-boson, where $T_3$ is the diagonal generator of $SU(2)_L$,
$Q_{\rm em}$ is the electromagnetic charge, and $\theta_W$ is the weak mixing angle.
Note that the neutrino's charge under the neutral current $t$-channel exchange has already been globally factored out
in Eq.~(\ref{eq:dcs}).  In the case that the neutrino scatters off of an electron, there is an additional interference diagram
for the charged-current $t$-channel exchange of a $W$-boson where the final $(\nu_e,e^-)$ states are associated with
crossed vertices.  An analogous charged current interference diagram likewise exists
for anti-neutrino scattering off of an electron, although it manifests instead via the $s$-channel.
The functional form of Eq.~(\ref{eq:dcs}) is unchanged by these additions, which may be absorbed
into a simple redefinition of the scattering charges. Intuitively, the $W$ exchange is pure left,
adding a unit shift to each of $(g_v,g_a)$.
However, if the scattering source should happen to be of the antineutrino variety, as is the case for
a nuclear reactor, then the entire effective axial charge inherits a relative negative phase,
subsequent to conditional application of the positive unit shift for matched flavor.
Intuitively, under $CP$, the inversion $(L \Leftrightarrow R)$ flips the sign of $g_a$.
These transformations are summarized following.
\be
[g_v,g_a] \,\Rightarrow\, [(g_v+\delta_{X,e}),\,\pm\,(g_a +\delta_{X,e})]
\label{eq:qgoestoA}
\ee
There is an additional complication for nuclear scattering, insomuch as one must account for the embedding of quarks into nucleons,
which are in turn bound in the nucleus.  This is a situation familiar from dark matter physics, with embedding prescriptions given
for example in \cite{DelNobile:2013sia,Agrawal:2010fh,Dienes:2013xya,Hill:2014yxa}.
The vector couplings to protons and neutrons are respectively $g_v^p = 1/2 - 2{\rm{sin}}^2\theta_W$ and $g_v^n = -1/2$.
The value of $\theta_W$ has been measured at energies of $\sim$10MeV in atomic
parity violation experiments \cite{Porsev:2009pr}, at energies of $\sim1-8$MeV from reactor experiments \cite{Canas:2016vxp},
and has also been calculated in the $\overline{\rm{MS}}$ scheme at low energies \cite{Erler:2004in}. In the present work we
adopt the value ${\rm{sin}^2\theta_W} = 0.2387$. Given this value, one finds that the vector coupling to protons is
subdominant compared to that for the neutrons. The axial couplings for protons and neutrons are
respectively $g_a^p = +1/2$ and $g_a^n = -1/2$, or effectively a negative of the prior for anti-neutrino scattering.
The total axial coupling to nuclei is proportional to the expectation value of the nucleon
spin content in the nucleus, $\langle S_p\rangle$ and $\langle S_n\rangle$. These expectation values vary from
element to element, with the values of interest in this work for germanium and silicon given by \cite{Klos:2013rwa}.
Since the vector coupling adds coherently while the axial charge couples (still coherently)
to the differential spin of embedded nucleons, the
dominant contribution to \cns will arise from vector coupling to the neutron count $N=A-Z$,
yielding a total cross section that scales at leading order as $N^2$. For the case of electron scattering, we take the free electron approximation, giving rise to sharp features at the atomic energy levels.

\section{\cns in Simplified Model extensions of the Standard Model\label{sct:extensions}}

\begin{figure*}[ht]
\begin{tabular}{cc}
\includegraphics[height=5cm]{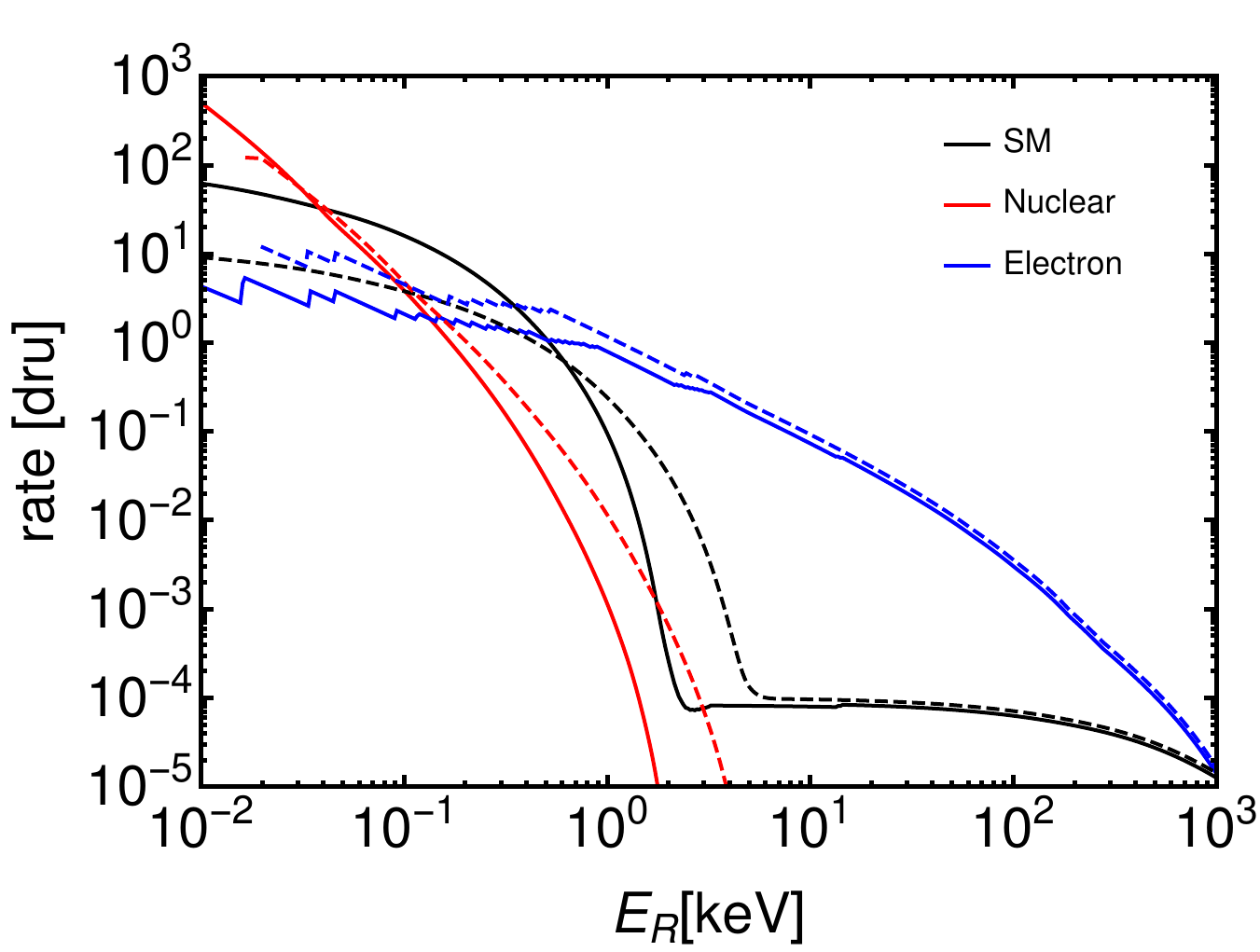} &
\includegraphics[height=5cm]{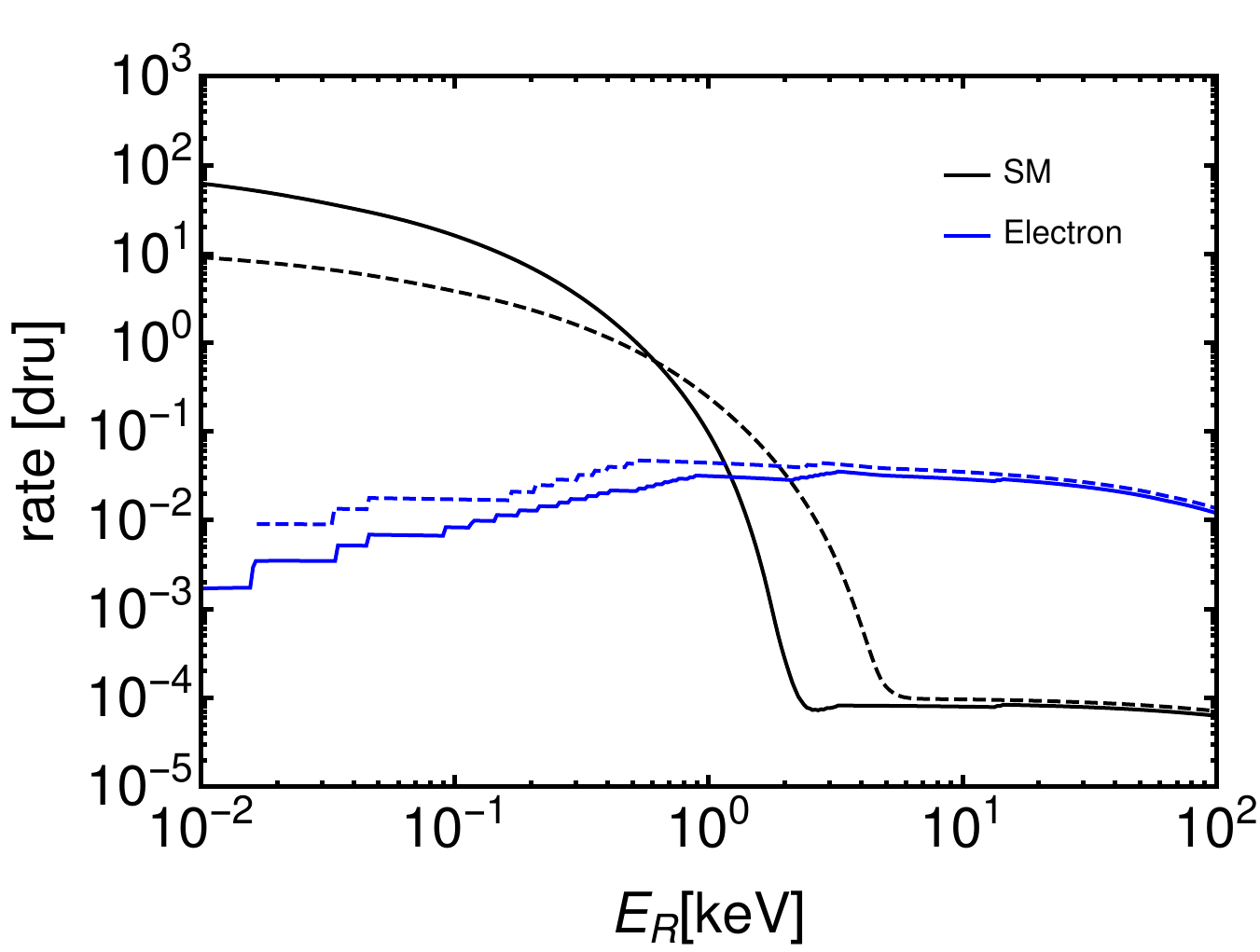} \\
\includegraphics[height=5cm]{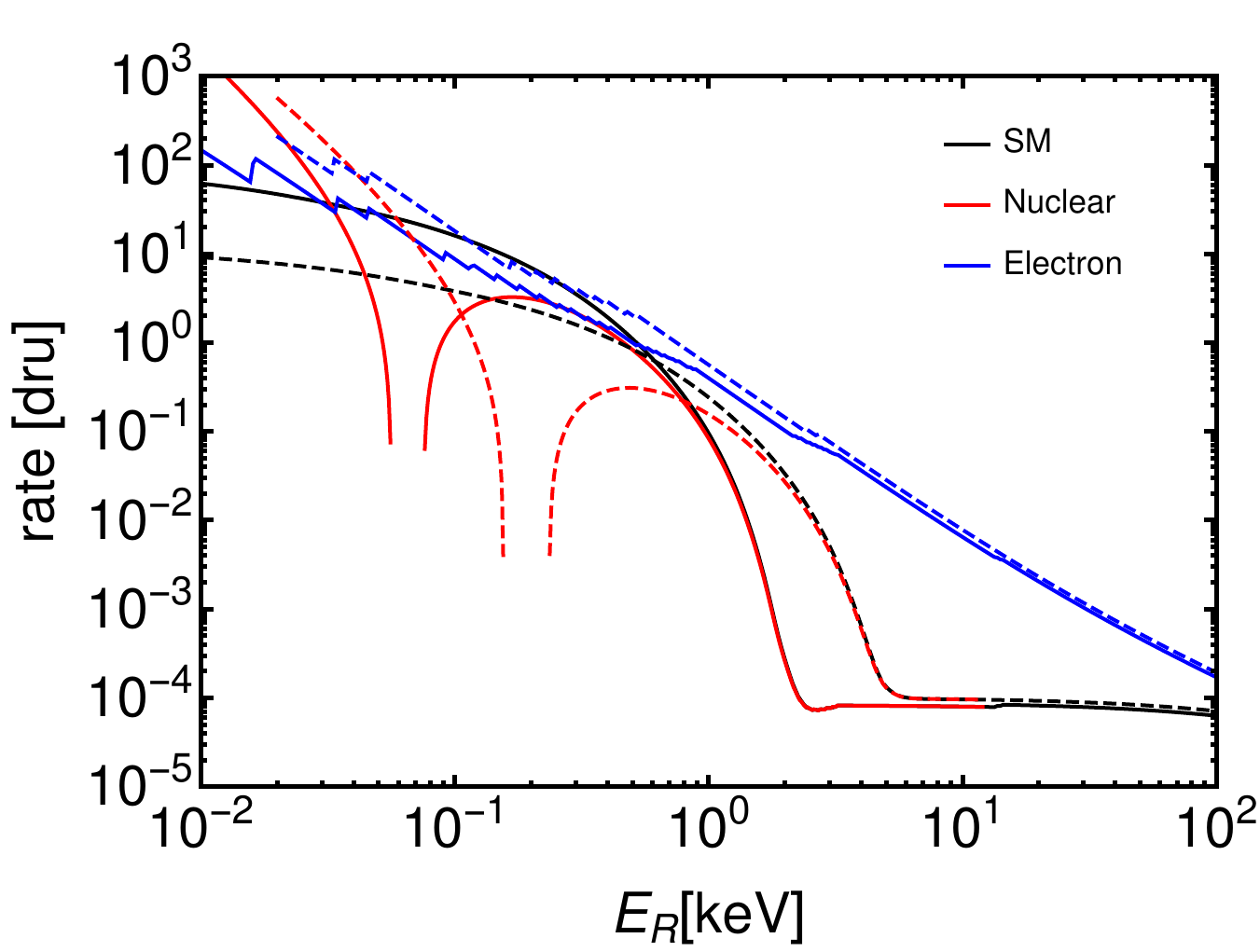} &
\includegraphics[height=5cm]{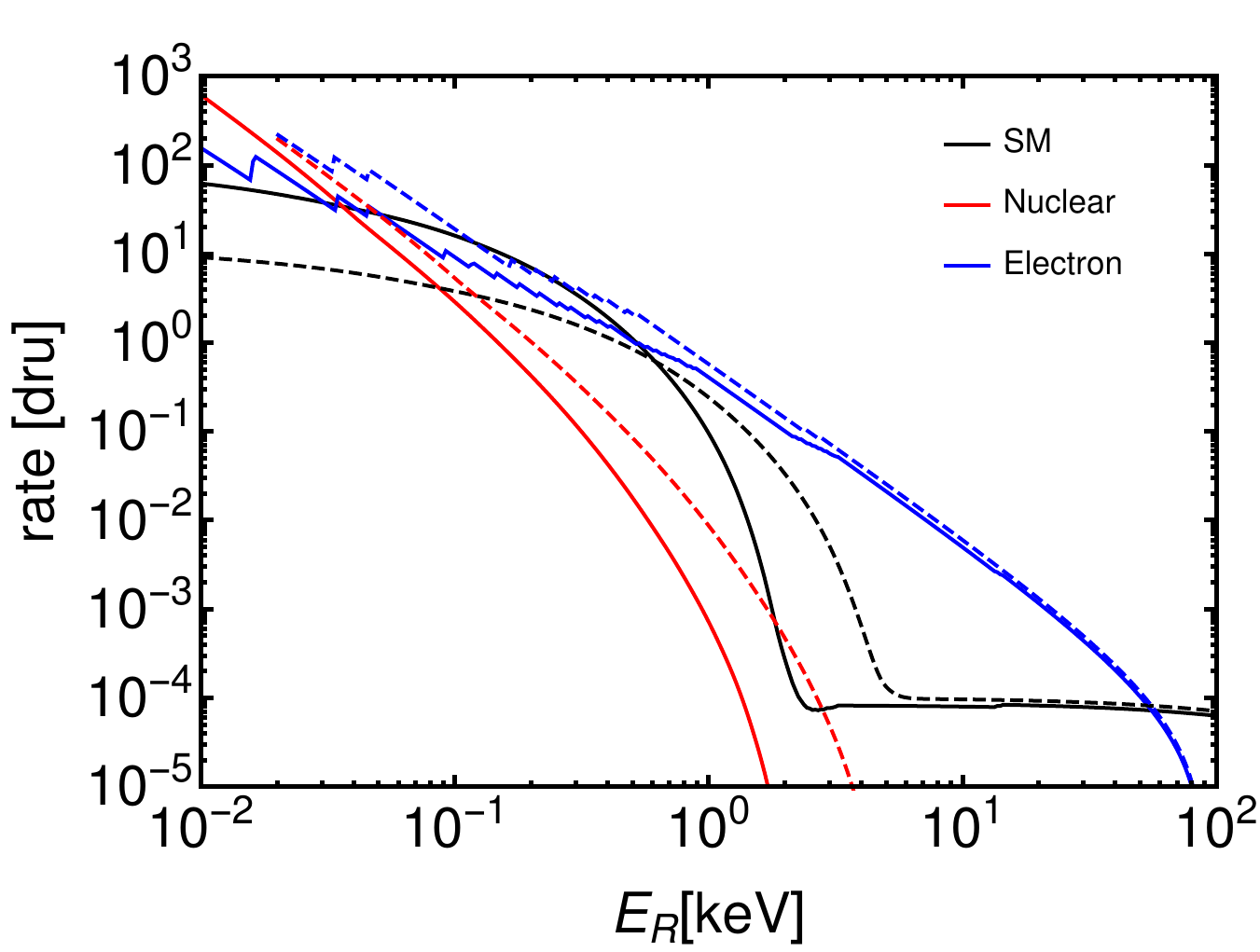} \\
\end{tabular}
\caption{Sample recoil rates for the four BSM scenarios with the following mediator types: scalar (top left),
pseudoscalar (top right), vector (bottom left), or axial-vector (bottom right). 
The solid and dotted lines are for Ge and Si detectors respectively. The mediator mass is taken to be 1keV and the coupling chosen such that the interaction is approximately the strength of the SM.
}
\label{fig:rates}
\end{figure*}

We next examine modifications to the leading \cns event profile induced by the presence of new neutral current light mediators.
For this analysis we will turn to a simplified model framework, which is a minimal extension of the SM that includes a new
mediating particle, along with its couplings to neutrinos, electrons, and quarks.  We follow a path similar to
that of \cite{Cerdeno:2016sfi}, where neutral current mediators are introduced that transform as a scalar (S),
pseudoscalar (PS), vector (V), or axial-vector (AV) under the Lorentz group.
The new interactions are specified by the following Lagrangians.
\bea
\mathcal{L}_{S} &&\supset \phi(g_{\nu,S}\bar{\nu}\nu+g_{\ell,S}\bar{\ell}\ell + g_{q,S}\bar{q}q)\\
\mathcal{L}_{PS} &&\supset \phi(g_{\nu,S}\bar{\nu}\nu-i\gamma^5(g_{\ell,PS}\bar{\ell}\ell + g_{q,PS}\bar{q}q))\\\
\mathcal{L}_{V} &&\supset Z_{\mu}'(g_{\nu,Z'}\bar{\nu}_L\gamma^{\mu}\nu_L + g_{\ell,v}\bar{\ell}\gamma^{\mu}\ell + g_{q,v}\bar{q}\gamma^{\mu}q)\\
\mathcal{L}_{AV} &&\supset Z_{\mu}'(g_{\nu,Z'}\bar{\nu}_L\gamma^{\mu}\nu_L - g_{\ell,v}\bar{\ell}\gamma^{\mu}\gamma^5\ell - g_{q,v}\bar{q}\gamma^{\mu}\gamma^5q)
\eea
Light mediators can be experimentally allowed, if their coupling is very weak.
In place of a four-point Fermi interaction vertex, there is in this case a product of
three-point couplings over a propagator explicitly dependent upon the momentum transfer $t = -2E_R m$.
A light vector or axial vector mediator may be simply accommodated in a manner similar to the prior inclusion
of charged current interference, via redefinition of the effective charges in Eq.~(\ref{eq:dcs}).
Defining the axial phase $\alpha$ of a target species $X$
by $[g_{X,v},g_{X,a}] \equiv g_{X,Z'}\! \times [\cos\alpha,\sin\alpha]$, the
prescription for holistic inclusion of these processes is
given by the following extension of Eq.~(\ref{eq:qgoestoA}).
\be
[g_v,g_a] \,\Rightarrow\, [(g_v+\delta_{X,e}),\,\pm\,(g_a +\delta_{X,e})]
+ \frac{g_{\nu,Z'}\, g_{X,Z'}}{\sqrt{2}G_F\, (2 E_R\, m_X + M_{Z'}^2)}
\times [\cos \alpha, \,\pm\,\sin \alpha]
\label{eq:qgoestoB}
\ee
As before, $(-)$ signs apply for the case of antineutrino scattering.
Note that the transformation in Eq.~(\ref{eq:qgoestoB}) will correctly reproduce the corresponding entries
in Table~IV of Ref.~\cite{Cerdeno:2016sfi} when
applied to the base neutral current cross-section of Eq.~(\ref{eq:dcs}),
while also automatically generating all suppressed terms of higher order in $E_R/E_\nu$.
For nuclear scattering, where $(E_\nu \ll m_N)$, we are likewise guaranteed
that $(E_R \ll E_\nu)$ will indeed apply. On the other hand, for electron scattering,
with $(E_\nu \simeq m_e)$, it may similarly be that $E_R \simeq E_\nu$.
However, investigation of the factor $2 E_R m_X + M_{Z'}^2$ from the
$Z'$ propagator suggests that the region of parameter space where
terms of higher order in $E_R/E_\nu$ become relevant is typically disjoint from
the region wherein the most substantial rate enhancement is derived from a light $Z'$ mediator.
Note first that the maximal kinetic recoil (for a head-on impact) impartable to a target of
mass $m_X$ is
\be
\tmax = \frac{2E_\nu^2}{m_X+2E_\nu}\,\,,
\label{eq:cutoff}
\ee
which reduces to the order of $m_e$ for $(E_\nu \simeq m_e)$.
In this case, $2E_R m_e$ cannot be much greater than roughly $(m_e^2 \approx E_\nu^2)$, but it can be much, much smaller.
This will be irrelevant if $(M_{Z'} \gtrsim m_e)$, such that the mediator mass-square will always dominate,
generating a flat response in $E_R$. However, if $(M_{Z'} \ll m_e)$, then the event rate is substantially
enhanced as $E_R$ decreases, scaling as $E_R^{-2}$ (taking the square of the amplitude),
until leveling off for recoils much below the knee $E_{R,e}^{\rm knee} \equiv M_{Z'}^2/2m_e$.
In particular, note that this cutoff for the rise in the light mediator enhancement occurs, given $(M_{Z'} \ll m_e \simeq E_\nu)$,
in a regime where $E_R \ll E_\nu$. No such similar statement may be made about the
neglect of terms higher order in $E_R/E_\nu$ if $(M_{Z'} \gtrsim m_e)$.
For the nuclear case, \tmax reduces to $2E_\nu^2/m_N$ for $(E_\nu \ll m_N)$.
As such, the relevant upper bound is $(2 E_R m_N \le 4 E_\nu^2)$. Correspondingly, there
will be no substantial variation of the light-mediator enhancement with $E_R$ unless
$(M_{Z'} \ll E_\nu)$.  If this is the case, then the event rate can again grow as
$E_R^{-2}$ until leveling off below an energy around $E_{R,N}^{\rm knee} \equiv M_{Z'}^2/2m_N$.
Note that even in the flat recoil response regime,
which may onset more or less concurrently for both nuclear and electron scattering
when $(M_{Z'}^2 \gtrsim 2\tmax\,m_X \approx E_\nu^2)$,
it remains possible, in principle, for the BSM rate to exceed the SM rate globally
if $(M_{Z'}^2 \lesssim g_{\nu,Z'} g_{X,Z'}/G_F)$.
All of the event features described here are purely kinematic, and are
broadly independent of the specific couplings or spin of the hypothetical mediator.

There are two important bits of conventional wisdom developed in the approach to coherent nuclear
scattering with heavy mediators that are now seen to be inapplicable, or at least not strictly applicable,
to the case of scattering mediated by a light field. The first is that any non-standard interaction (NSI)
must have cross-terms with the standard model in order to yield an appreciable rate.
The usual intuition here (cf.~Ref.~\cite{Dutta:2015vwa}) is that NSI
vertices, e.g.~those representing a heavy gauge boson $(M_{Z'} \gg M_Z)$ with SM-comparable coupling strengths,
will necessarily have amplitudes of substantially smaller magnitude than those of the SM, such that the
leading new physics contribution in the square of summed amplitudes is from the interference term.
A corollary of this statement is that new physics diagrams that cannot interfere with the SM,
e.g. those representing flavor-changing neutral currents, will be practically invisible to experiments
that are not sensitive to the final state on an event-by-event basis. For light mediators, it is
quite conceivable that the failure of prior experiments to observe the corresponding signatures of
new physics has been purely one of technology, namely the inability to probe very soft recoils of the target.
Given a sufficient advance in technology to render the final state visible, it need not actually be weak,
and a boost from interference with known SM processes is thus no longer a prerequisite to discovery.
The second element of standard lore to be upended in the present context is that
electron scattering is always sub-dominant to nuclear scattering in visibility at low recoil momenta.
Support for this observation comes from two directions, the first being that only the nucleus is
able to remain coherent for deBroglie wavelengths typical of nuclear energies,
and thereby access the $\mathcal{O}(10^2)$ relative rate enhancement
associated with summing over constituents in the amplitude prior to squaring.
The second advantage (or sometimes disadvantage) of nuclear scattering is purely kinematic.
The cutoff scale for recoil scattering \tmax of a heavy nucleus may typically be a factor
of $10^4$ larger than the corresponding cutoff for electron recoil.  Consequently, the nuclear
recoil spectrum is highly concentrated inside a very narrow energy bandwidth, wherein its relative
amplitude is correspondingly elevated.  If the resultant recoils falls below threshold sensitivity
of the detector, then they will be invisible.  However, if the necessary threshold can be breached,
a substantial gain in rate is attained.  Whereas competing backgrounds to detection may have a
very wide spectral composition, this isolation of a highly targeted integration domain is quite beneficial.
Additionally, although event-by-event discrimination of nuclear from electron recoils is typically
lost in solid state detectors at very soft recoils (as the phonon vs. ionization yield curves converge
or the underlying phonon signal is traded for threshold sensitivity via Luke gain, cf. CDMSlite~\cite{Agnese:2015nto})
a reasonable amount of population discrimination may be recovered simply from this radical disparity of profiles.

In the regime $(M_{Z'}^2 \ll E_\nu^2)$, with typical MeV scale solar or reactor neutrino energies,
both the electron and nuclear $Z'$ scattering amplitudes will grow as $1/E_R$ with decreasing recoil energy,
until flattening out at $(E_{R,X}^{\rm knee} \simeq M_{Z'}^2/2m_X)$.  In particular, the electron recoils will
both onset $(E_{R,e}^{\rm max}/E_{R,N}^{\rm max} \simeq 10^4)$,
and level off $(E_{R,e}^{\rm knee}/E_{R,N}^{\rm knee} \simeq 10^5)$,
at much larger recoils for the same $M_Z'$.
With identical couplings $g_{X,Z'}$, the single electron scattering
to nuclear scattering (per particle constituent) amplitude ratio will grow linearly with increasing $E_R$,
from parity below $E_{R,N}^{\rm knee}$, up to a ceiling around $10^5$ above $E_{R,e}^{\rm knee}$.
This advantage can, in principle, be more than sufficient to overcome the coherency advantage of the nucleus.
However, the extended slow decline of the electron scattering rate across regions of the
recoil parameter space well above several keV, which have historically been well instrumented and well probed,
implies that the corresponding coupling must be very weak.
Nevertheless, it may remain possible to skirt existing bounds while strongly enhancing the electron recoil
at very low energies.  For example, with $E_\nu = 2$~MeV, $M_{Z'}=1$~keV, and $g_{\nu,Z'} g_{e,Z'} = 10^{-13}$,
the new physics rate from electron scattering on a germanium target always exceeds the new rate from
nuclear scattering, drops below the SM rate above 10~keV, and sharply exceeds the SM nuclear coherent scattering
rate below about 15~eV, recovering the advantage of bandwidth compression conventionally held by the nuclear recoil.
Moreover, this contribution may still be disentangled from the standard SM coherent nuclear scattering
signal via identification of the characteristic $E_R^2$ power law enhancement shape.

Interestingly, the electron recoil differential cross section can grow more rapidly (in a logarithmic sense)
with increasing coupling than the nuclear recoil at low energies.  The reason for this is that the new physics
contribution will include both a cross term with the SM, and a square of the new physics amplitude, and
the SM contribution is already large for the nuclear scattering case.  When the new physics amplitude is
smaller than the SM amplitude, the scaling with BSM couplings will be linear, but when the new physics
amplitude is larger, the scaling with BSM couplings is quadratic.

The possibility of a scalar or pseudoscalar mediator $\phi$ has several clear distinctions from
that of a neutral vector mediator.  Most essentially, a scalar coupling to two spinors mixes
left and right chiralities, as would a mass term, whereas the vector mediator necessitates
the insertion of a vector of gamma matrices between the spinors, which will not mix chirality.
If it exists, the right-handed neutrino chirality has no standard model interactions.  Indeed, the
solar or reactor source will consist of solely left-handed neutrinos (or right-handed
anti-neutrinos).  However, the described experimental construction is
insensitive to the nature of the exiting neutrino, and there is no direct prohibition
against a sterile neutrino playing this role.  More importantly, however, given that the SM has
no process with identical initial and final states, there will be no SM interference terms
generated (again, these are not necessarily vital to prospects for detection).
Additionally, the scalar and pseudoscalar scenarios are prohibited from mixing,
such that only the squares of individual couplings are referenced in the differential cross section.
This has the additional interesting consequence that neutrino and anti-neutrino scattering are
equivalent, and insensitive to relative inversions of phase.  Additionally, incompatibility
of the Lorentz structure implies that it is impossible in this case to tidily summarize the new physics
as a shift in the couplings of the neutral current SM vector boson exchange.
The full additions to Eq.~(\ref{eq:dcs}), for a single particle with couplings
$[g_{X,S},g_{X,PS}] \equiv g_{X,\phi}\! \times [\cos\alpha_\phi,\sin\alpha_\phi]$ to the field $\phi$, are as follows.
\begin{equation}
\frac{d\sigma}{dE_R} \supset
\frac{g_{\nu,\phi}^2 g_{X,\phi}^2 (E_R^2 m_X + 2 E_R m_X^2 \cos\alpha_\phi)}{8\pi E_\nu^2 (2 E_R m_X + M_\phi^2)^2}
\label{eq:dcsscalar}
\end{equation}

In Fig.\ref{fig:rates} we have plotted example recoil spectra for $\bar{\nu}_e - $nucleus and $\bar{\nu}_e - e^-$
scattering in germanium and silicon detectors.  These spectra highlight
several important points: i) the motivation for low threshold detectors is obvious from the rise in the spectrum at
low recoil energies, ii) the new interactions can produce deviations from the SM that could be within reach for
near term experiments, and iii) the interference between the pure SM interaction and the SM plus simplified model
interactions may produce destructive interference effects from $V$ and $AV$ exchange.

\section{Ultra-low threshold detectors with reactor neutrinos\label{sct:detectors}}

In order to calculate the rate of recoil for nuclei we must specify the flux and energy distribution of
incident neutrinos. Here we primarily consider reactor neutrinos, detected via \cns using ultra-low threshold detectors. However, for comparison of sensitivity we will also consider Solar neutrinos and stopped pion sources. These flux components are summarized in table~\ref{tab:Flux}.

\begin{table}[tb]
\caption{Neutrino flux sources and their respective uncertainties in the flux normalizations. The SNS flux and uncertainty was taken from~\cite{Akimov:2015nza}. The Solar components are derived from the high metallicity Solar model as outlined in Ref.~\cite{Robertson:2012ib}.}
\begin{tabular}{c|c}
 component  & $\nu$ flux (cm$^{-2}$s$^{-1}$) \\
\hline
TAMU reactor (at 1m) & $1.50(1\pm0.02) \times 10^{12}$ \\
\hline
SNS (at 20m)         &   \\
$\nu_\mu$ (prompt) & $4.30(1\pm0.1) \times 10^{7}$ \\
$\nu_e$ (delayed)  & $4.30(1\pm0.1) \times 10^{7}$ \\
$\bar\nu_{\mu}$ (delayed) & $4.30(1\pm0.1) \times 10^{7}$ \\
\hline
Solar        &    \\
pp           & $5.98(1\pm0.006) \times 10^{10}$ \\
$^7$Be       & $5.00(1\pm 0.07) \times 10^9$ \\
$^8$B        & $5.58(1 \pm 0.14) \times 10^6$ \\
pep & $1.44(1 \pm 0.012) \times 10^8$ \\
\hline
\end{tabular}
\label{tab:Flux}
\end{table}

To investigate the reach of reactor neutrino sources in probing NSI with
light mediators, we consider the Mitchell Institute Neutrino Experiment at a Reactor (MINER). As a brief review,
the MINER program has been developed with the Nuclear Science Center at Texas A\&M University (TAMU),
which administrates a megawatt-class TRIGA-type pool reactor stocked with low enriched ($\sim20\%) {}^{235}$U.
Low temperature solid state germanium and silicon detectors,
using technology similar to that currently developed for
direct dark matter searches like SuperCDMS,
will be installed at very near proximity ($\sim1-3$m) to the reactor core.
More specifics on the reactor, its properties, and the MINER program
may be found in recent works which have highlighted its physics potential for TeV scale mass $Z'$ models,
sensitivity for neutrino magnetic moment searches, and sterile neutrino searches \cite{Dutta:2015vwa,Dutta:2015nlo}.

The antineutrino flux can be obtained via knowledge of the reactor's power (1.00$\pm$0.02 MW in the present work)
along with the normalized antineutrino fission spectrum, which has been measured at various sites (for a recent
discussion of the current status of the spectrum see \cite{An:2016srz}).
The spectrum has not been directly measured below the 1.8 MeV inverse beta decay threshold.
For these energies we adopt the theoretical distribution
in Ref.~\cite{Kopeikin:2012zz}.
Above 1.8 MeV we use the experimental results of Ref.~\cite{Schreckenbach:1985ep}.

We assume a detector exposure corresponding to germanium and silicon masses each of 10kg,
together with a five year running time, as is realistic for the MINER experiment.  As previously
mentioned, an extremely important aspect of the detector technology is the existence of low thresholds for nuclear
recoils (nr), due to the expected features in \cns that arise at recoil energies below the keV scale. We
assume a 100eVnr threshold for germanium and silicon. While no current
technology exists that reaches such thresholds, experimental progress is underway which can conceivably reach
these levels in the near future \cite{Mirabolfathi:2015pha}.  For example, CDMSlite recently achieved a threshold of 56eVee (electron recoil) \cite{Agnese:2015nto} and a few hundred eVnr (the exact conversion factor from eVee to eVnr
depends on the parameterization for the ionization yield, which is only known to within a factor of a few).
The proximity to the reactor, while ensuring a large neutrino flux, comes with a commensurately large neutron and gamma background. The goal for the MINER experiment is
approximately 100 events per day per kg per keV, i.e. 100 dru, in the signal region. Modeling of the reactor, detector and shielding show that this goal should be achievable~\cite{Agnolet:2016zir}, and thus we will use a flat background of 100dru as a baseline.
These detector specifications are summarized in table~\ref{tab:Detectors}. In addition, we include a more
optimistic future scenario (Ge/Si II), to show the improvement that detector technology developments could yield.

\begin{table}[ht]
\caption{Detector specifications}
\begin{tabular}{r|r|r|r|r|r|r|r|}
Name & Target & Exposure (kg.days) & $E_{th}$ (eV) & background (dru) \\
\hline
Ge & germanium & 10,000  & 100  & 100$\pm$10 \\
Ge II & germanium & 10,000 & 10  & 10 $\pm$1 \\
Ge II(low BG) & germanium & 10,000 & 10  & (1$\pm$.1)$\times10^{-4}$ \\
Si & silicon & 10,000 & 100  & 100$\pm$10 \\
Si II & silicon & 10,000 & 20  & 10 $\pm$1 \\
CsI   & Caesium-Iodide & 10,000 & 5,000 & 10$\pm$1 \\
\hline
\end{tabular}
\label{tab:Detectors}
\end{table}

The distribution of solar neutrinos presents either large flux
at low energies, or low flux at large energies, as in table~\ref{tab:Flux}.  The highest flux rates are from the pp
process, integrating to approximately $6\times10^{10}~{\rm cm}^{-2}{\rm s}^{-1}$ and cutting
off around 0.4~MeV, followed by the ${}^7{\rm Be}$ and pep line sources at 0.9 and 1.4~MeV
with corresponding fluxes $5\times10^9$ and $1\times10^8~{\rm cm}^{-2}{\rm s}^{-1}$
(largely dominant over the N, O, F continuum), and the broad ${}^8{\rm B}$ source that extends out
to more than 15~MeV with an integrated flux of $6\times10^6~{\rm cm}^{-2}{\rm s}^{-1}$.
In order to access the \cns process with near-term detector technology, it is
necessary that the neutrino source have an energy in the few MeV range.  For example,
a nuclear recoil detector threshold around (20, 10)~eV is required in (silicon, germanium)
in order to register about half of the scattering events from neutrinos with a mean energy of
1.5~MeV~\cite{Dutta:2015vwa}.  The threshold scales as a square with $E_\nu$, such that
$E_\nu = 0.5$~MeV would require sensitivity at the level of about (5, 2.5)~eV, approaching
the theoretical limit for single electron ionization resolution~\cite{Mirabolfathi:2015pha}.
Electron recoils require no such dramatic sensitivity, but they generally suffer with respect
to rate and diffuse background contamination for the reasons described previously.
For the spectral components where \cns could be presently visible, the flux is woefully low.
By comparison, a nuclear reactor presents an (electron antineutrino) spectral composition
distributed across the few MeV range that can exceed all of the solar flux by orders
of magnitude, typically $10^{12-13}~{\rm cm}^{-2}{\rm s}^{-1}$~\cite{Singh:2003ep,Wong:2005vg,Dutta:2015vwa}.
This combination presents obvious advantages for the rate of signal events, although it
likewise presents new challenges with respect to the management
of radiologically intense backgrounds~\cite{Agnolet:2016zir}.

In addition to neutrinos from reactor and solar sources, stopped pion sources, such as at the SNS, produce neutrinos through the decay of muons at rest. These sources present the lowest fluxes, but at a much higher energy with a well known flux. The higher energy allows the use of more conventional detectors with modest thresholds of a few keV, such as the CsI detectors employed in the first phase of the COHERENT experiment~\cite{Akimov:2015nza}.

\section{Computation of confidence intervals and limits\label{sct:smconflimits}}

\begin{figure*}[ht]
\begin{tabular}{cc}
\includegraphics[height=5cm]{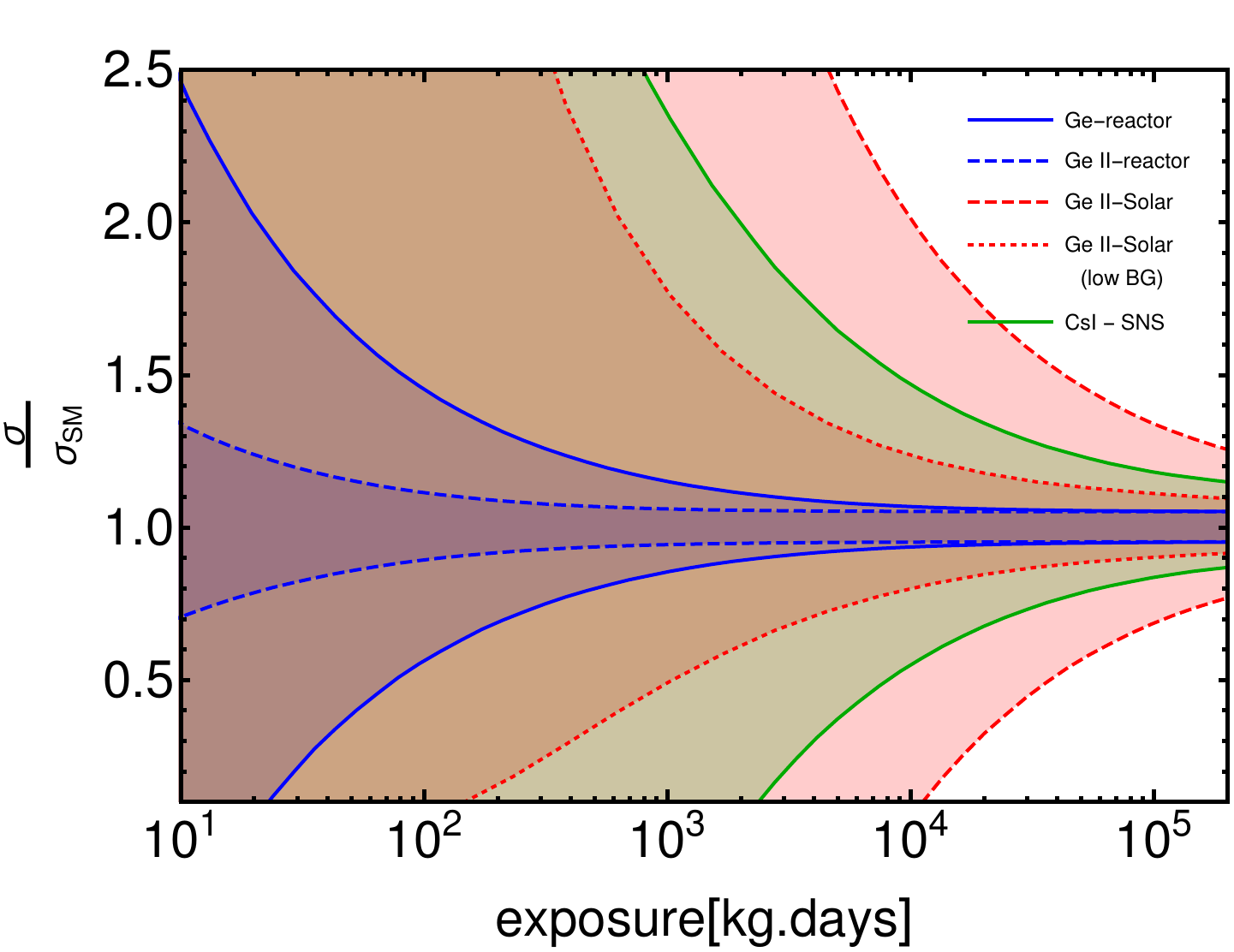} &
\includegraphics[height=5cm]{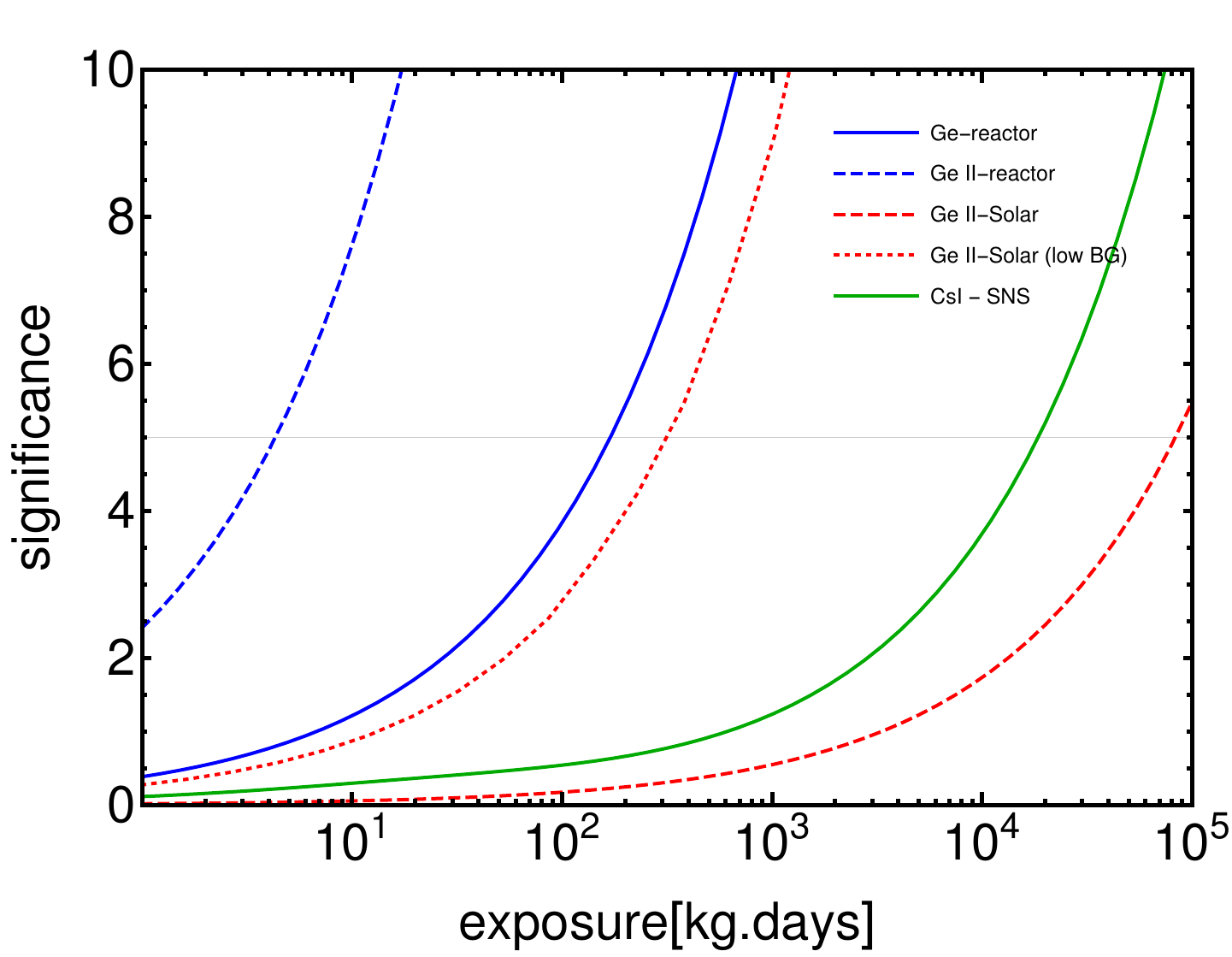} \\
\end{tabular}
\caption{\textit{Left:} 90\% confidence intervals on the standard model \cns cross section from a selection of experiments. \textit{Right:} exposure required to reach a 5-$\sigma$ fluctuation from the background only hypothesis.}
\label{fig:SMconfInt}
\end{figure*}

Before competitive limits can be placed on new physics, the standard model \cns process must be discovered and measured.
The MINER experiment can expect $\sim10$ dru from SM \cns, providing a signal-to-noise ratio of 1:10. From this we estimate
that a 5$\sigma$ signal will be observed with a few months of run-time. To quantify how well the \cns cross section can be
measured we will use the profile likelihood test statistic:
\be
t_\mu = -2 \mathrm{log}  \frac{\mathcal{L}(\mu,\hat\theta)}{\mathcal{L}(\hat\mu,\hat\theta)},
\label{eq:proflike}
\ee
where $\mu$ is the signal strength, $\frac{\sigma}{\sigma_{\mathrm{SM}}}$, and $\theta$ represents the nuisance parameters.
Hatted parameters denote a maximization. We will use a binned likelihood function given by:
\be
\mathcal{L} = \prod_i \frac{\nu_i^{n_i} e^{-\nu_i}}{n_i!} \prod_{j} e^{-\frac{1}{2\delta_j^2}(1-N_j)^2}.
\label{eq:binnedlike}
\ee
Here $\nu_i$ and $n_i$ are the expected (SM) and observed events in each bin, and the second product is a Gaussian
likelihood summed over the nuisance parameters: the background and flux component
normalizations $N_j$ (one, three and four components for
the reactor, SNS and solar cases respectively, where the values are given in table \ref{tab:Flux}).
The test statistic is then used to derive the expected 90\% confidence intervals on the
signal strength using the Asimov dataset~\cite{Cowan:2010js}.
Fig.~\ref{fig:SMconfInt} shows the confidence intervals and discovery significance as a function of exposure for several example experiments.
The MINER experiment (situated at 2m from the reactor) will be able to discover the SM process within 100kg$\cdot$days and
constrain the SM cross section to $\pm10\%$ with 10$^3$kg$\cdot$days of exposure, even with the conservative
background and threshold assumptions. 
The solar and SNS neutrino experiments require roughly 100 times as much exposure to achieve the same discovery potential and measurement accuracy. This is due to the smaller fluxes and larger uncertainties in these experiments.

To investigate the experimental reach of the MINER experiment we calculate discovery limits for the
detector configurations listed in table \ref{tab:Detectors}, based on the flux of the TAMU reactor at 2m.
The limits are also calculated for the same detectors with Solar neutrinos as the source.
For comparison with a current experiment we have also included the discovery limits for the first phase of the COHERENT experiment. The
various neutrino sources cannot be distinguished by the detector, and thus the reactor and SNS cases should thus include the solar neutrinos
as well, although the limits are displayed separately for illustration.  Given that a dedicated solar experiment would likely be carried out deep
underground (not in close proximity to a nuclear reactor), we have also included the calculation for a Ge II
experiment in a background-free environment.  The discovery limits are defined as the smallest signal
that could produce a 3$\sigma$ fluctuation 90\% of the time. To find this limit we use a binned likelihood
function, Eq.~\ref{eq:binnedlike} with log-spaced bins. The likelihood is used to calculate the log-likelihood ratio
and generate the test statistic $q_0$, from which we may estimate the expected median significance
of an experiment via the Asimov dataset~\cite{Cowan:2010js}.
\be
q_0 =
\begin{cases}
   -2 \mathrm{log}  \frac{\mathcal{L}(\mu=0,\hat\theta)}{\mathcal{L}(\hat\mu,\hat{\hat\theta})}  & \sigma  \geq \hat\sigma \\
   0	& \sigma < \hat\sigma \\
\end{cases}
\label{eq:loglike}
\ee

\begin{figure*}[ht]
\begin{tabular}{cc}
\includegraphics[height=5cm]{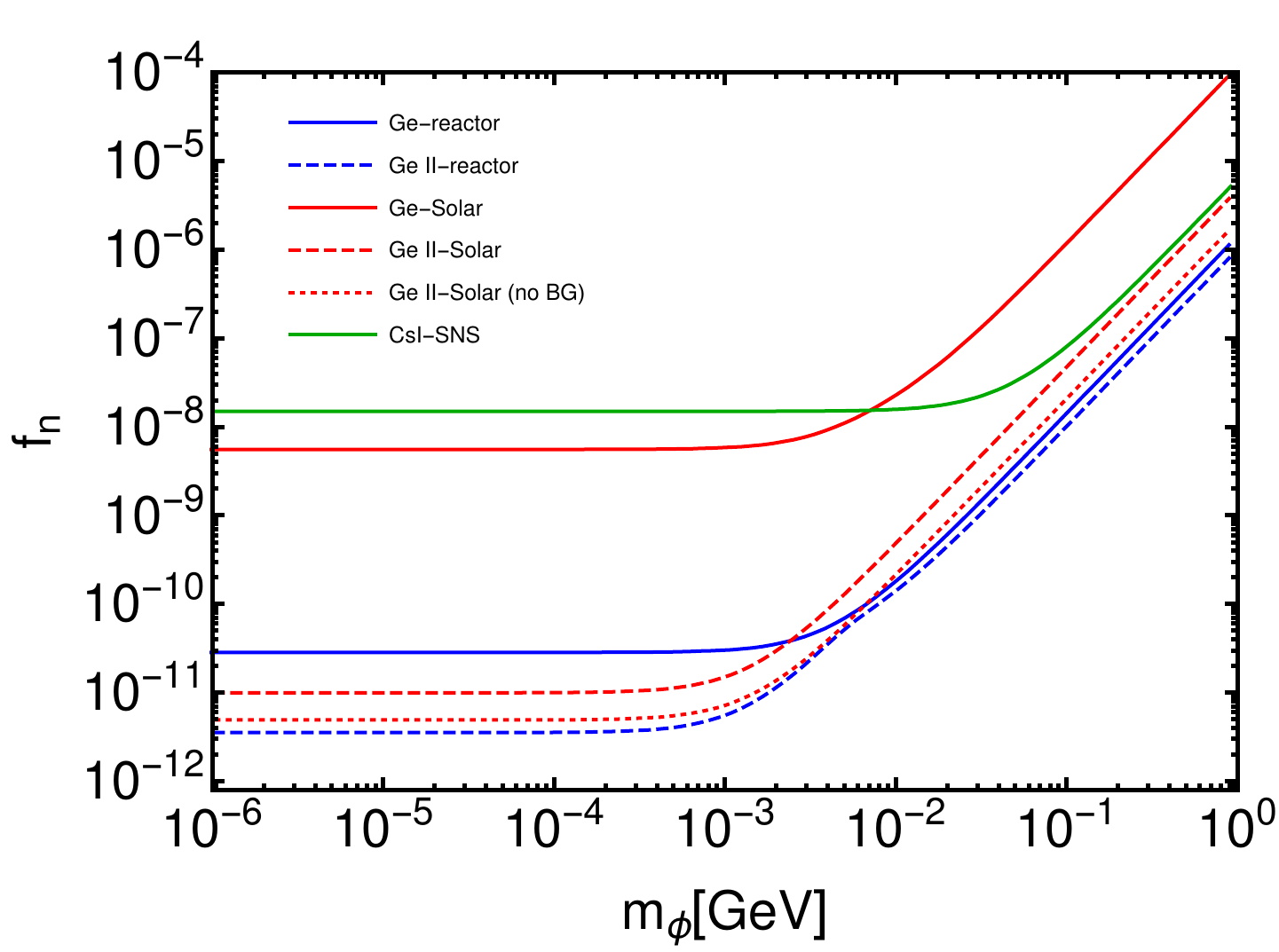}  &
\includegraphics[height=5cm]{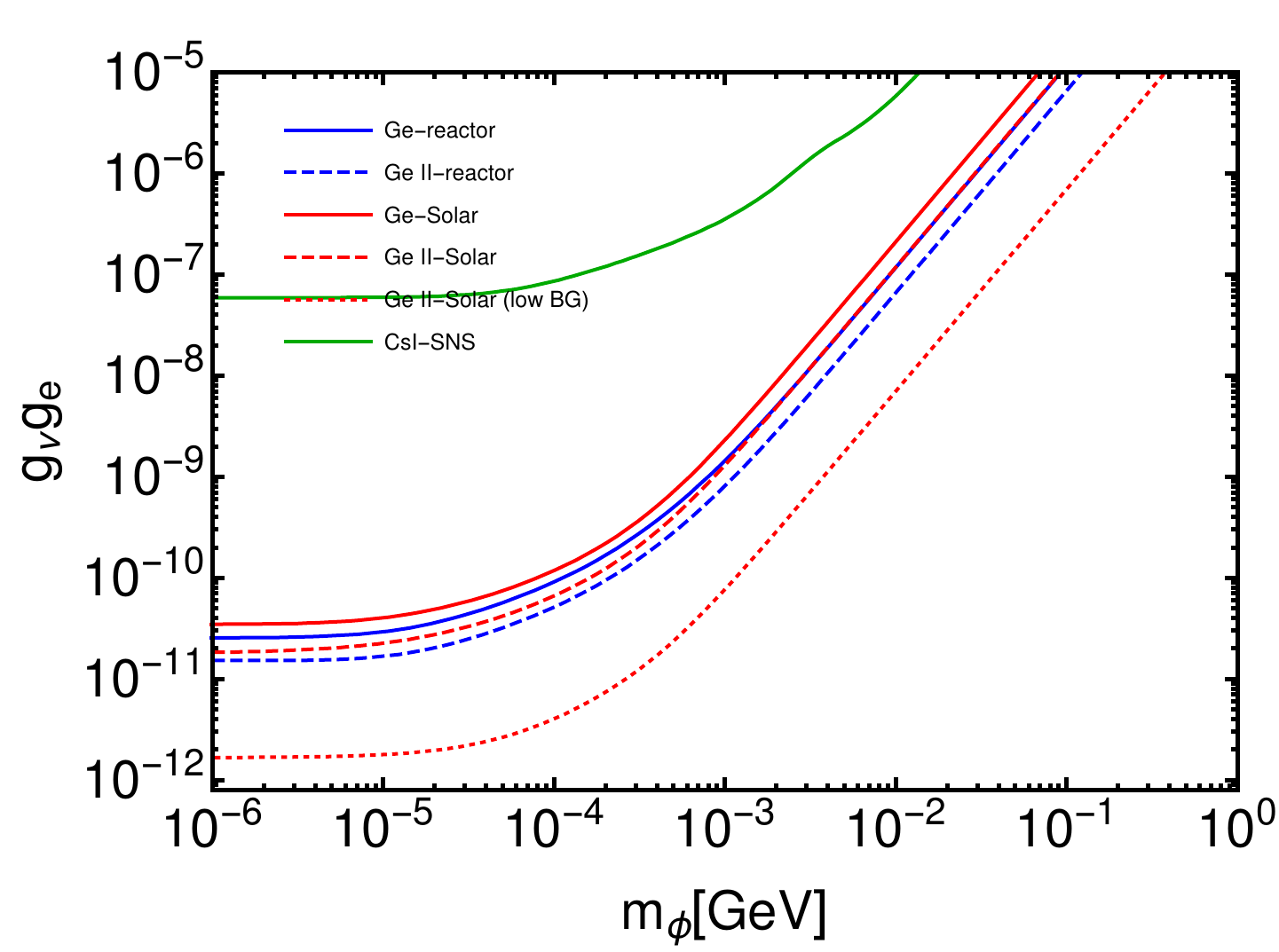} \\
 &
\includegraphics[height=5cm]{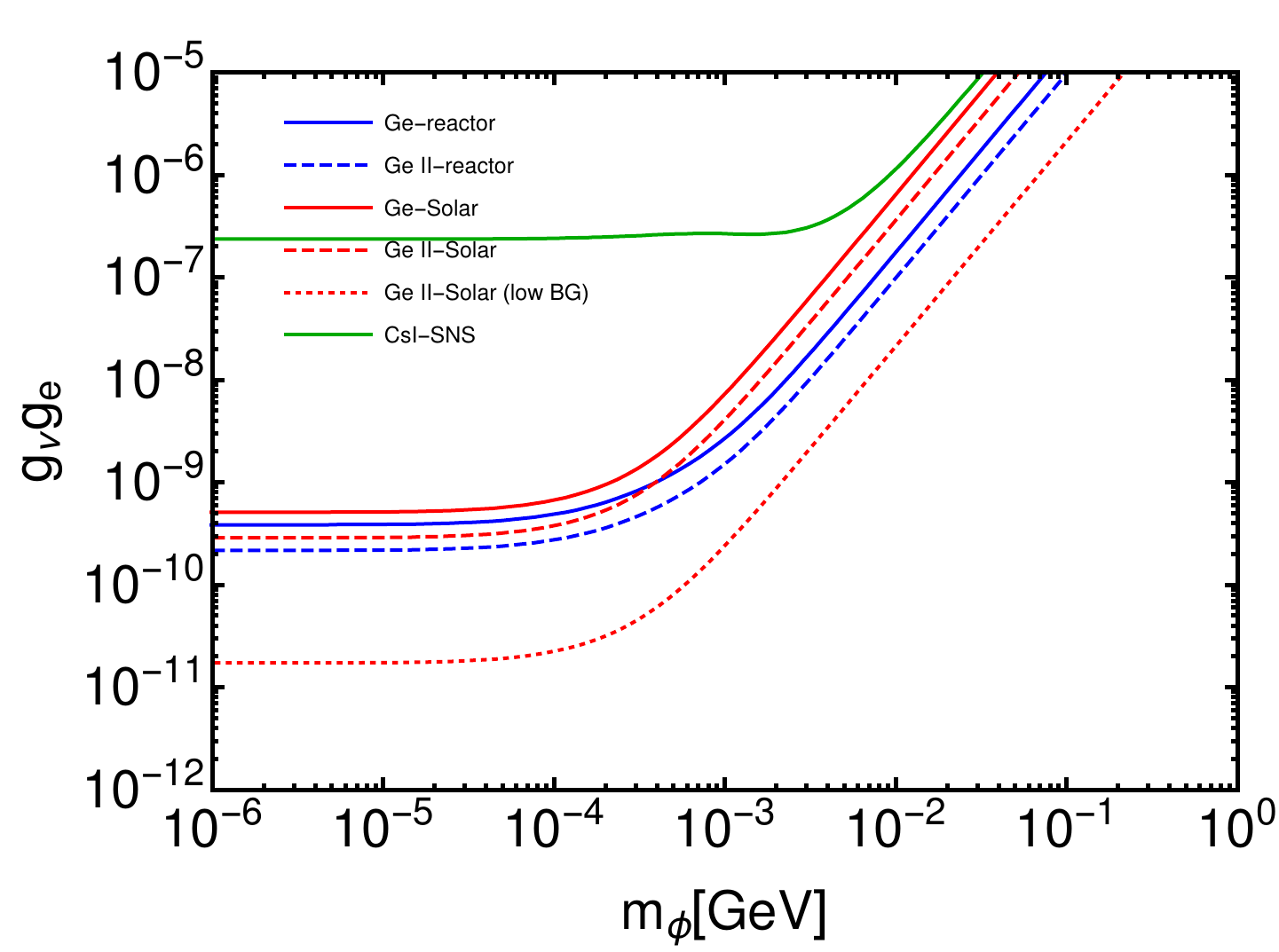} \\
\includegraphics[height=5cm]{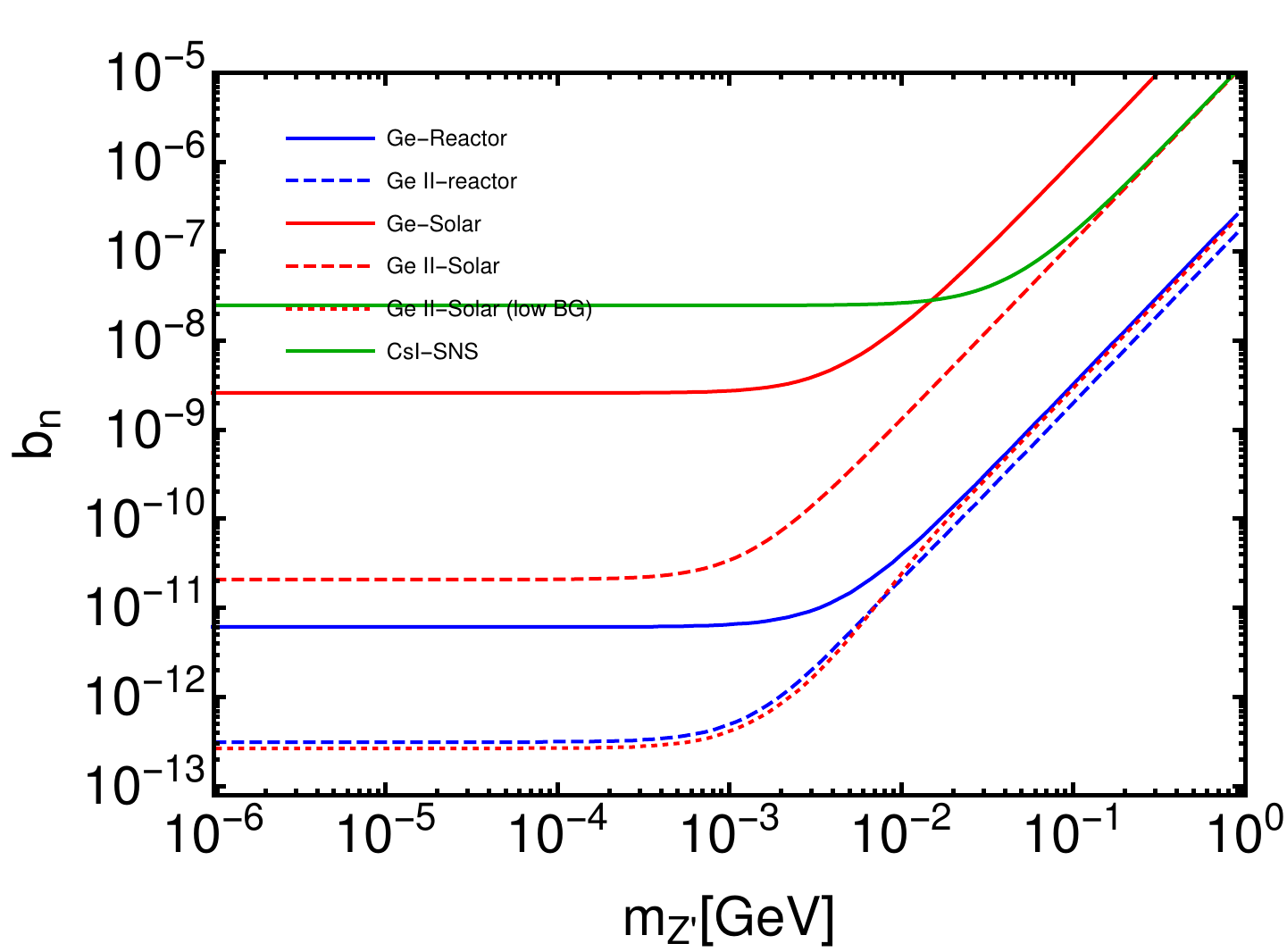}  &
\includegraphics[height=5cm]{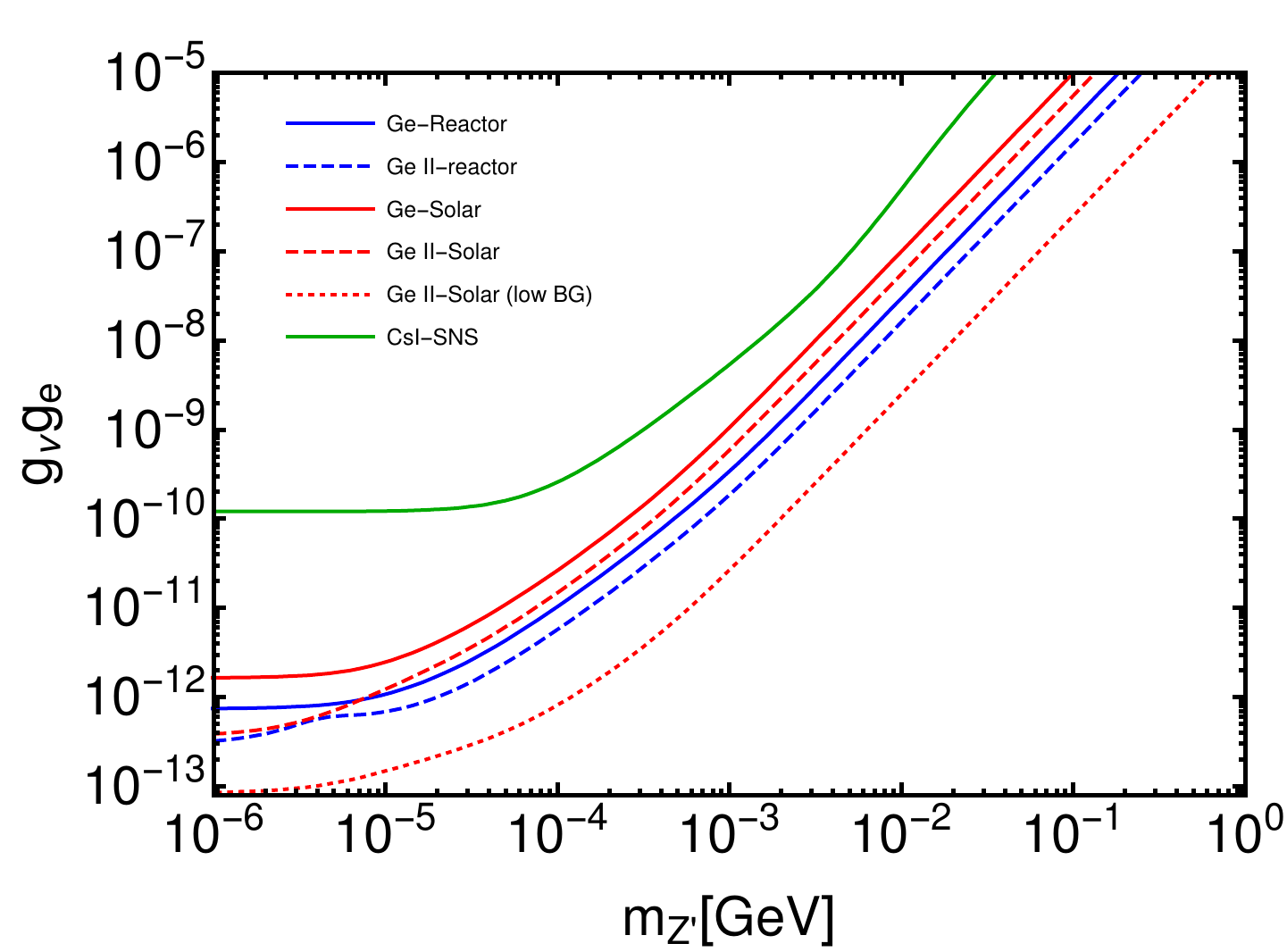} \\
\includegraphics[height=5cm]{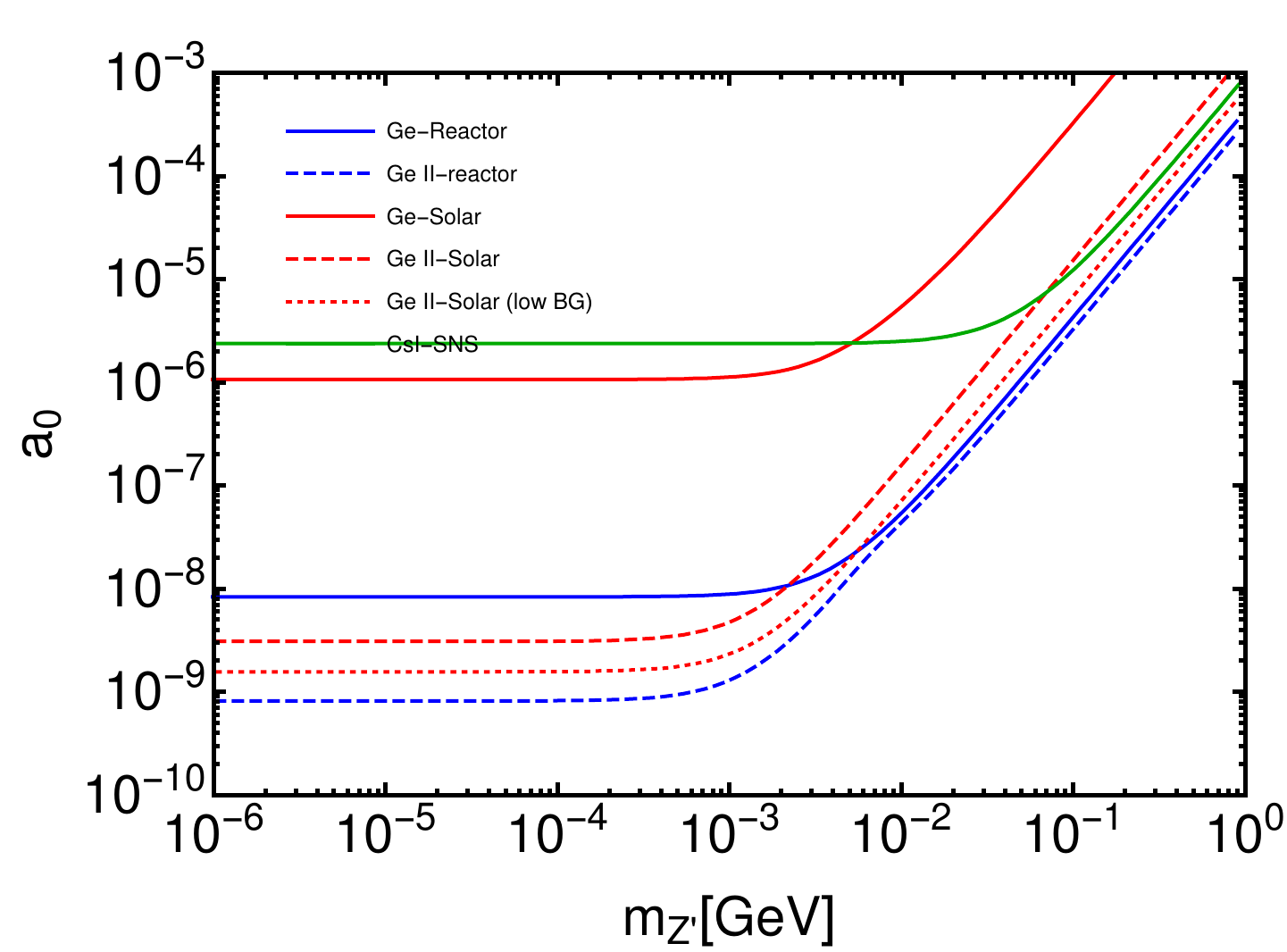}  &
\includegraphics[height=5cm]{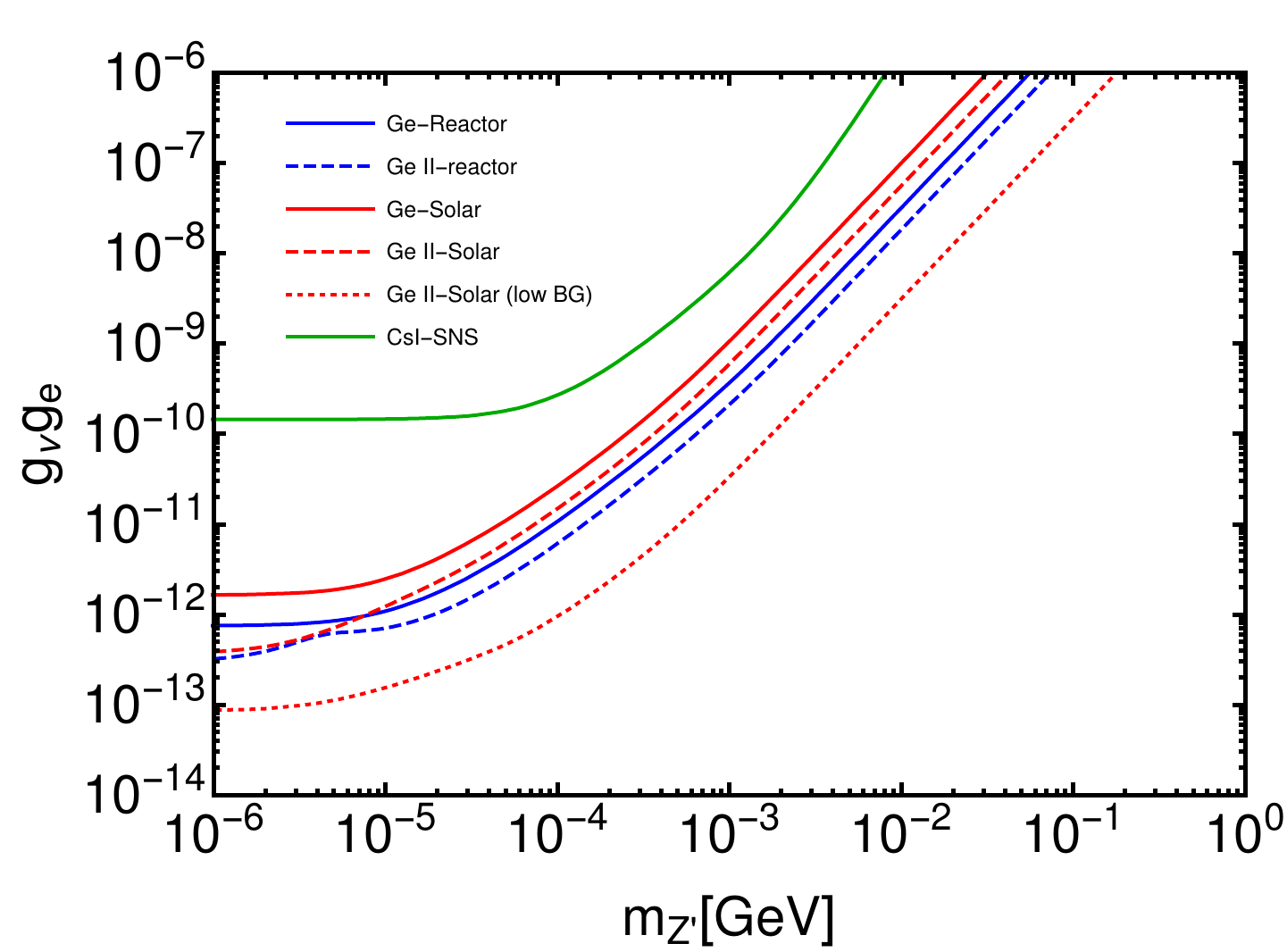} \\
\end{tabular}
\caption{Discovery limits for neutrino scattering off germanium nuclei (left) and electrons (right),
for the different BSM models (from top to bottom): scalar, pseudo-scalar, vector and axial-vector }
\label{fig:discLimits}
\end{figure*}

\begin{figure*}[ht]
\begin{tabular}{cc}
\includegraphics[height=5cm]{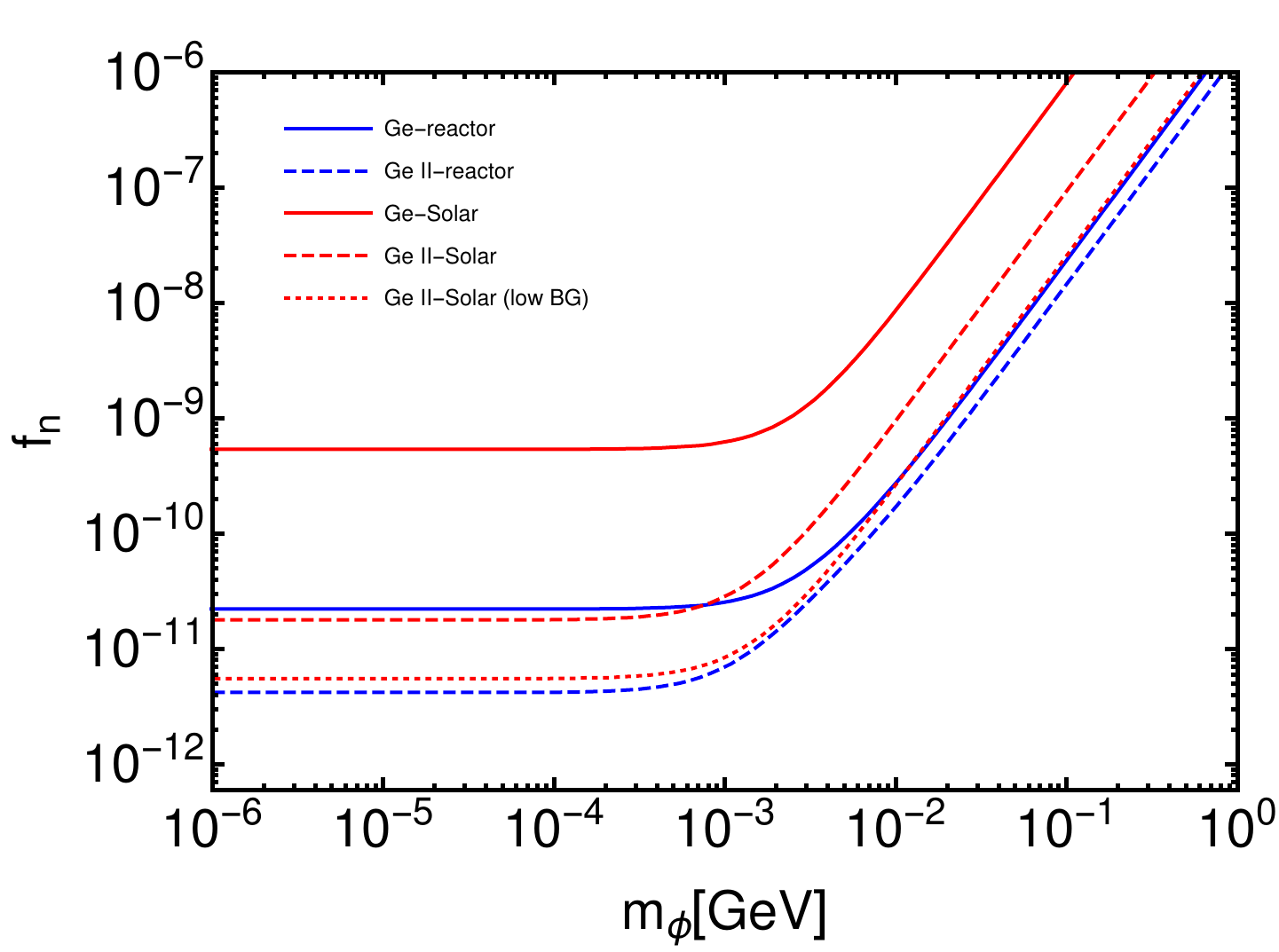}  &
\includegraphics[height=5cm]{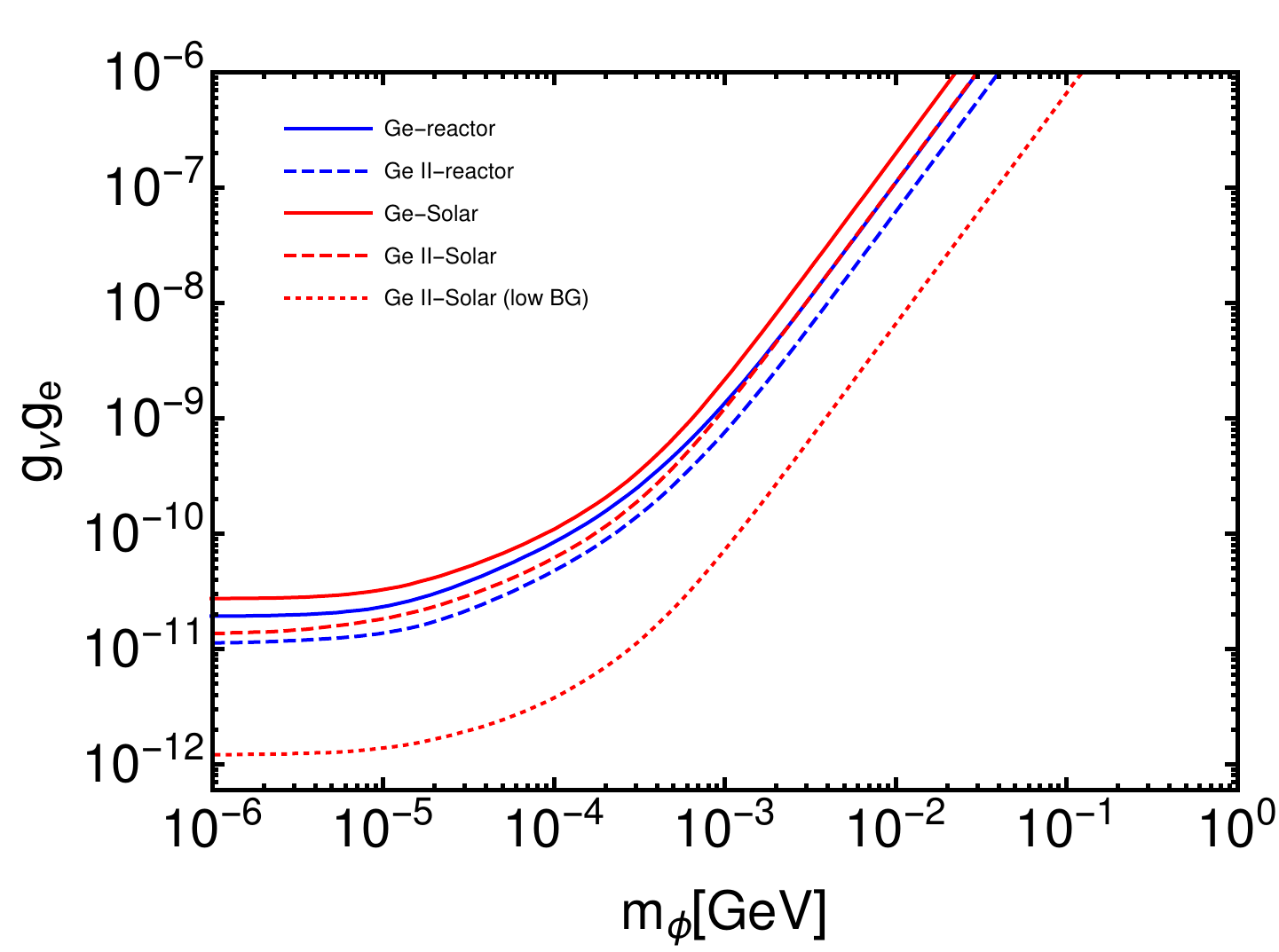} \\
 &
\includegraphics[height=5cm]{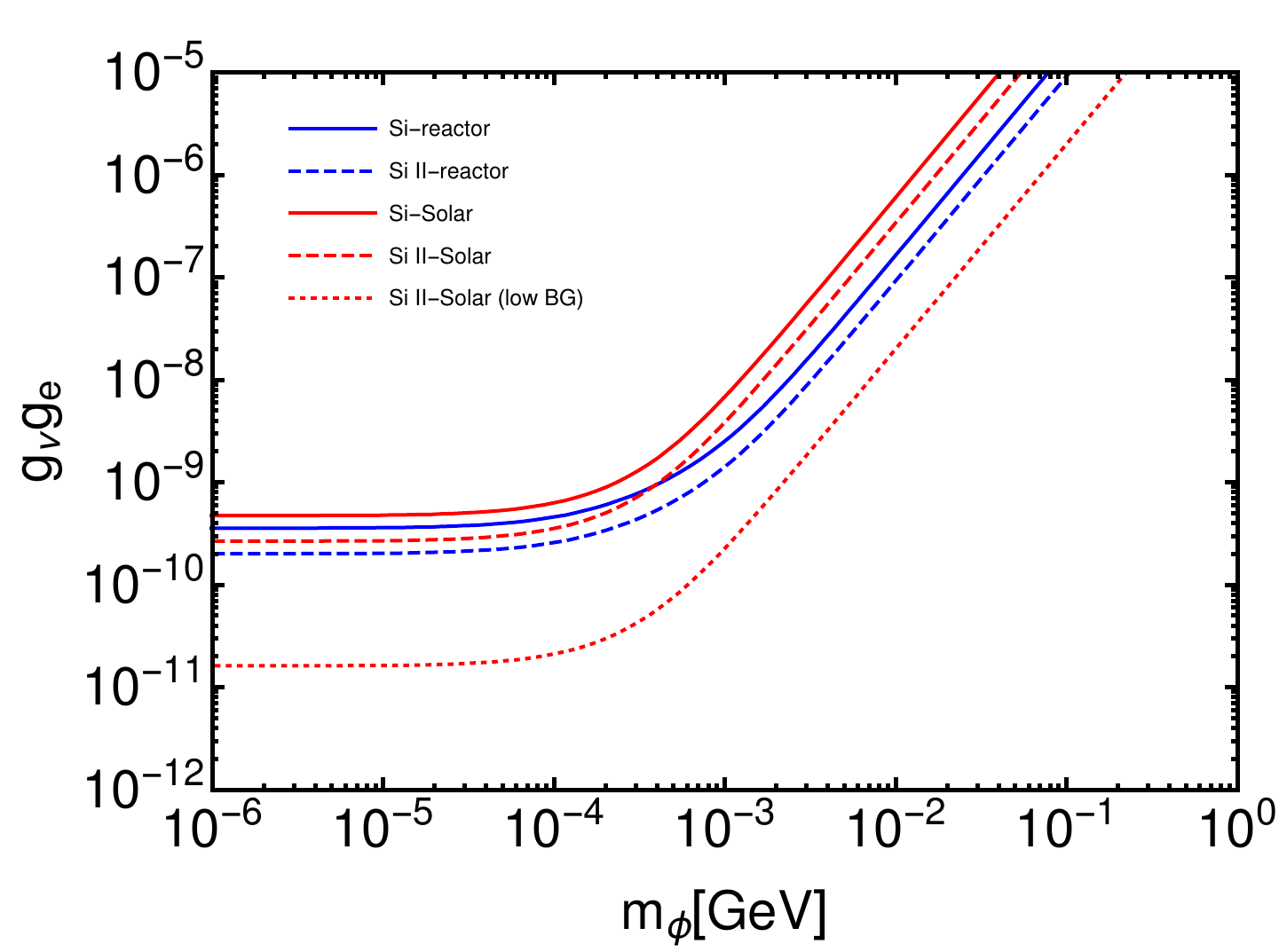} \\
\includegraphics[height=5cm]{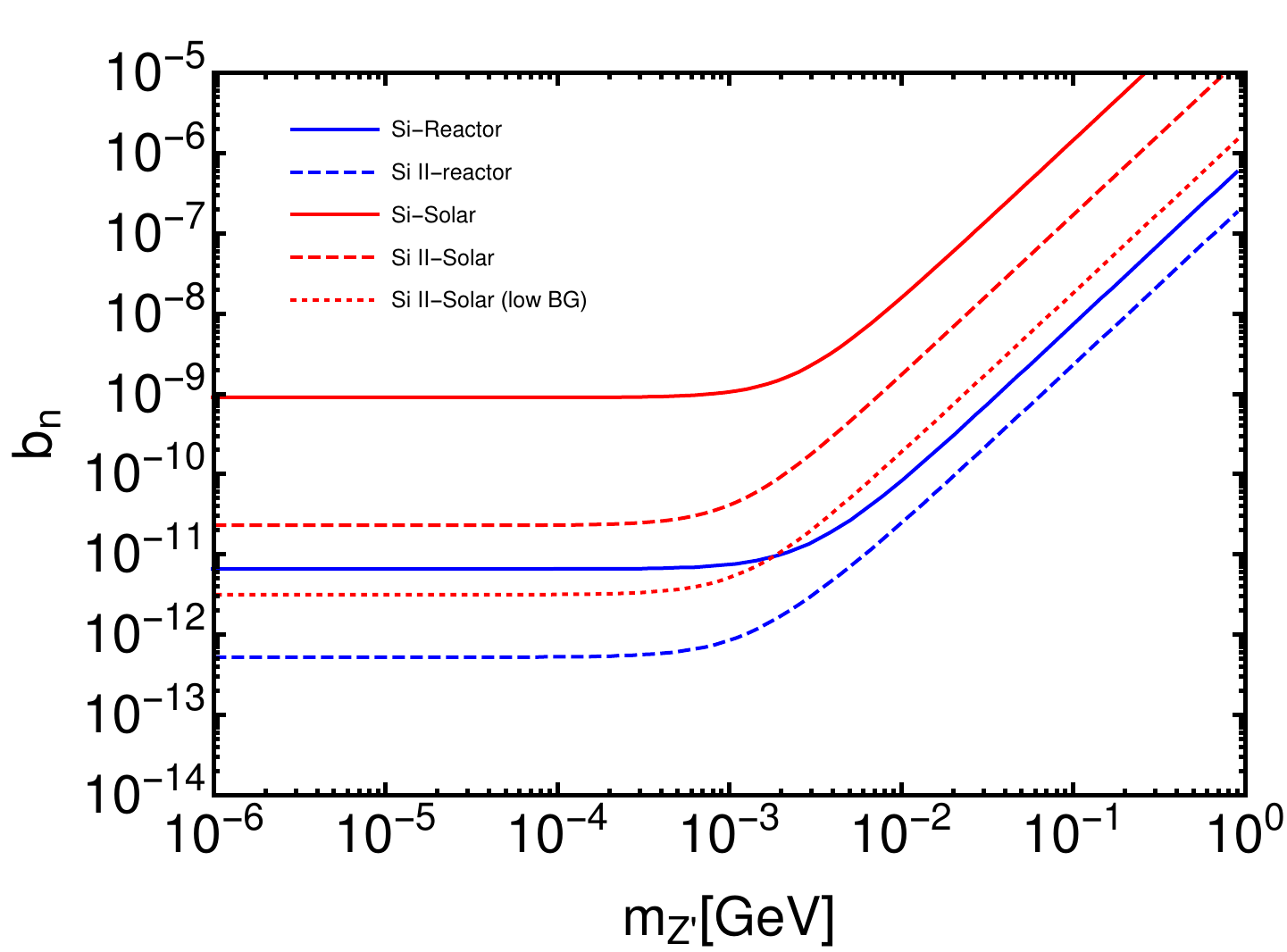}  &
\includegraphics[height=5cm]{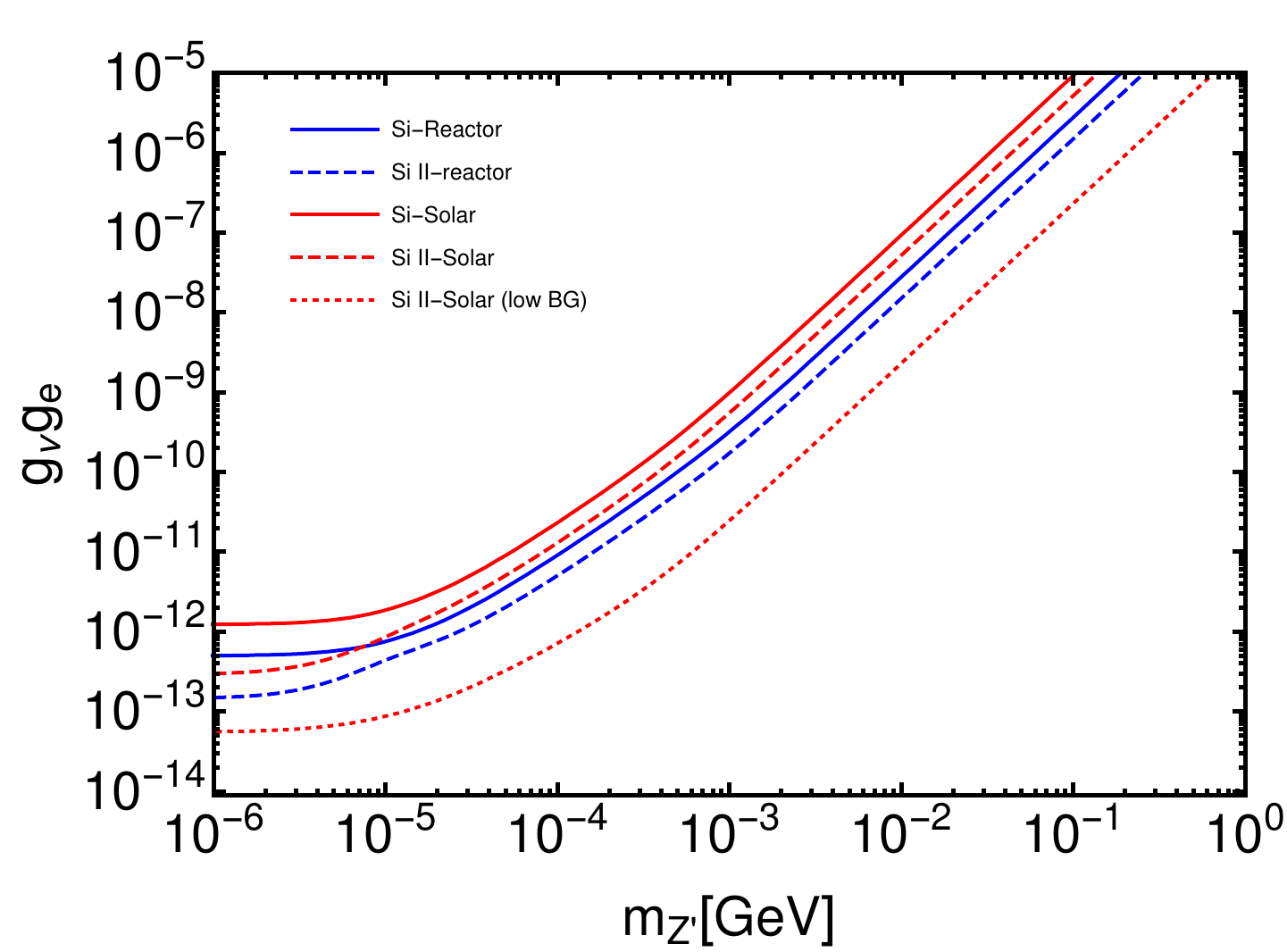} \\
\includegraphics[height=5cm]{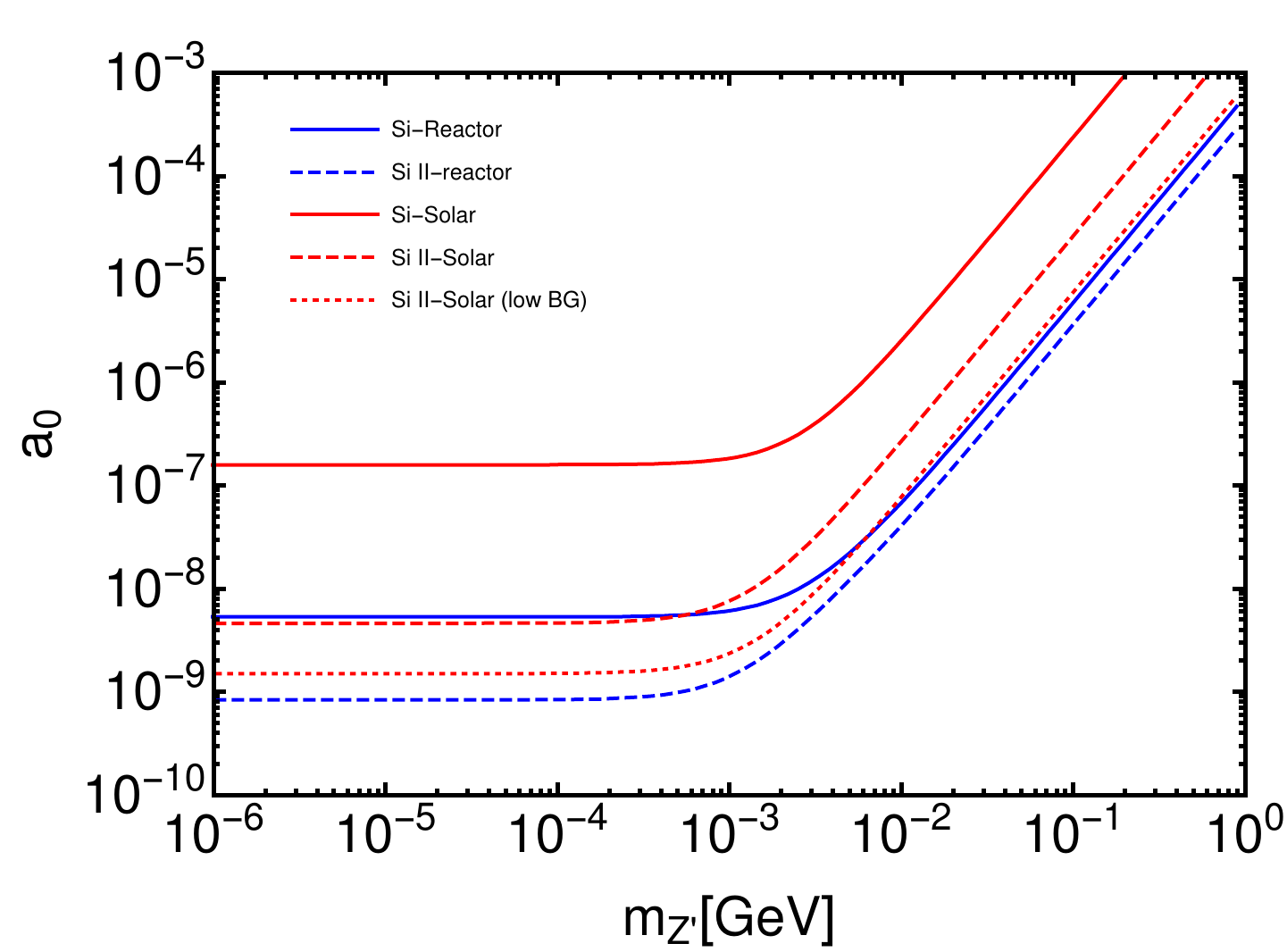}  &
\includegraphics[height=5cm]{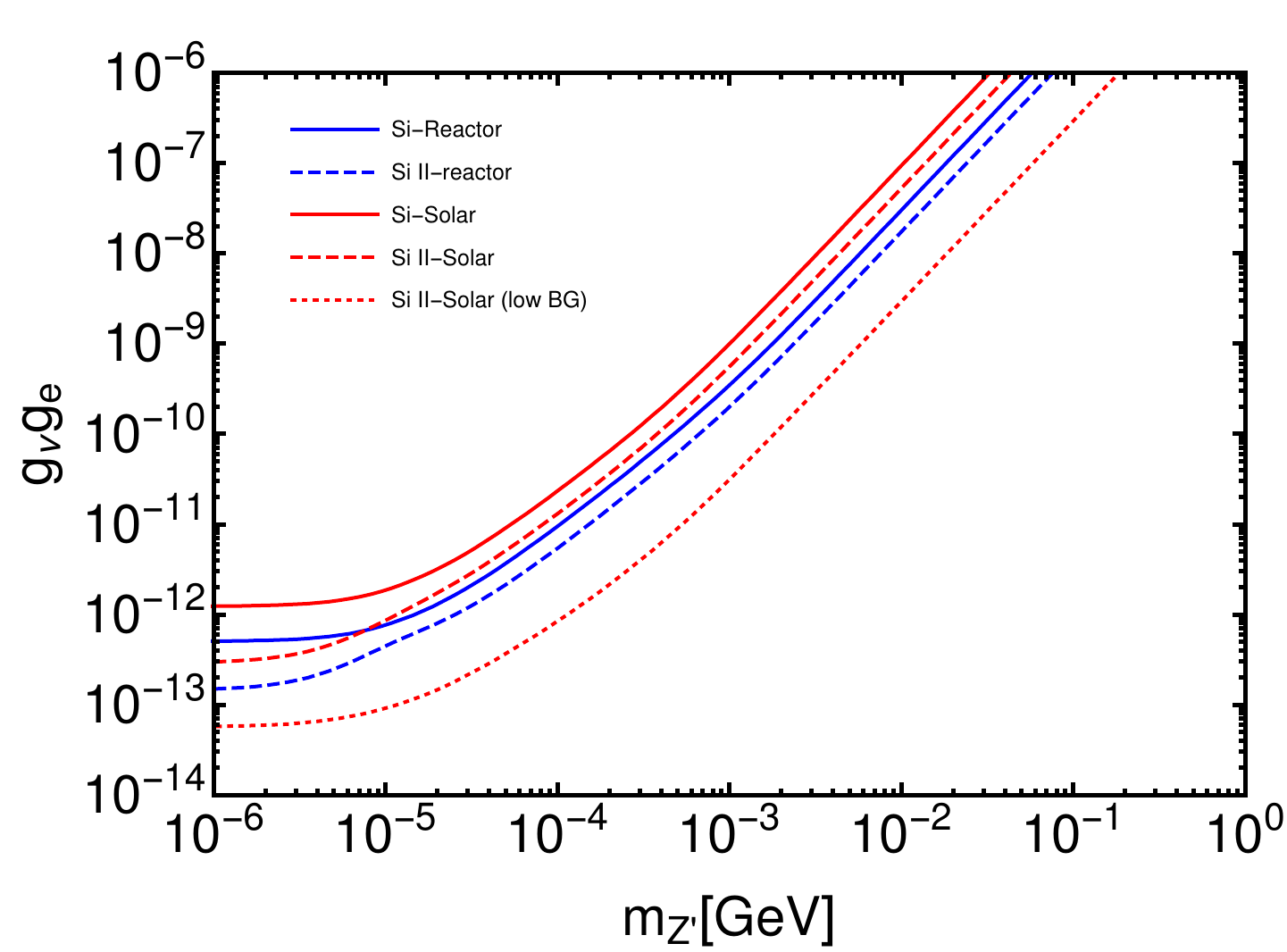} \\
\end{tabular}
\caption{The same as Fig.~\ref{fig:discLimits} but for silicon detectors}
\label{fig:discLimitsSI}
\end{figure*}

\section{Discovery limit results\label{sct:limits}}

The discovery limits for each of the four BSM models are shown in figure~\ref{fig:discLimits}.
For nuclear scattering, the parameters $f_n$, $b_n$ and $a_0$ shown in the plots are
respectively the scalar, vector, and axial vector analogs of
the electron scattering coupling product $(g_{\nu,Z'}\, g_{e,Z'})$,
but at the per nucleon or per unit spin level.
We assume no isospin violation, such that the corresponding coherency factors
for the nucleus at large are simply $(Q'_s = f_n A)$, $(Q'_v = b_n A)$, and $(Q'_a = a_0 \langle S \rangle)$.
A first glance through each of the Fig.~\ref{fig:discLimits} plots will reveal
an essentially similar ``hockey stick'' curve profile, the general shape of which
can be understood in terms of the momentum transfer.

As the mediator mass is decreased, the cross section increases,
and thus the experiments are sensitive to smaller values of the couplings.
For heavier mediator masses, the coupling (or rather the product of mediator couplings to the
neutrino and to the scattering target) sensitivity is diminished with a log-log slope of 2, due to
effects of the new physics interaction and propagator in the large-mass (point-like interaction) regime,
where one has a Fermi styled contribution $g_{\nu,Z'} g_{X,Z'}/M_{Z'}^2$.
This ratio is constant along the $(\log\,g_{\nu,Z'} g_{X,Z'} = 2\log\,M_{Z'})$ line.
Conversely, the mediator mass may be neglected when it becomes
small compared to the momentum transfer.  The relevant term here is
$g_{\nu,Z'} g_{X,Z'}/2 E_R m_X$, which corresponds to a flat coupling limit sensitivity.
There is a transition regime that interpolates between these two asymptotes, wherein the coupling
sensitivity bends smoothly.
Looking again at the propagator, we can infer that
this knee in the sensitivity should be centered around $M_{Z'} \simeq \sqrt{2 E_R m_X}$.
This knee occurs at a lower mediator mass for electron scattering since the
momentum transfer is much smaller in that case.
Limiting momentarily to the dominant vector interaction of Eq.~(\ref{eq:dcs}),
and factoring out couplings and constant terms, the functional dependence of the
differential cross-section may be very well approximated as a linearly declining
function of $E_R$ that cuts off at \tmax, as specified in Eq.~(\ref{eq:cutoff}).
\be
\frac{d\sigma}{dE_R} \,\,\appropto\,\,
1-\frac{E_R}{\tmax}
\label{eq:dcsprop}
\ee
It should be understood that the light vector mediator and scalar mediator interactions
imply complications to this simplified structure,
as in Eq.~(\ref{eq:qgoestoB}) and Eq.~(\ref{eq:dcsscalar})
respectively, although certain universalities persist.
For a detector sensitivity threshold $\tmin$, and considering a monochromatic neutrino
source energy $E_\nu$, the integrated area under this curve is
$(\tmax-\tmin)^2/2\tmax$, or simply $\tmax/2$ if the threshold $(\tmin \ll \tmax)$
is much smaller than the cutoff.
Likewise, the mean recoil is $\langle E_R\rangle = (2\tmin+\tmax)/3$, and
the standard deviation of the recoil is $\sigma_{E_R} = (\tmax-\tmin)/3\sqrt{2}$.
This suggests that the sensitivity knee for nuclear recoils should be centered
around $E_\nu$, or around the geometric mean $\sqrt{m_e E_\nu}$ for electron recoils.
The propagated width in the mediator mass is
$\sigma_{M_{Z'}} = \sqrt{m_X/2\langle E_R\rangle}\times\sigma_{E_R}$,
which implies a fractional width
$\sigma_{M_{Z'}}/M_{Z'} \simeq 1/2\sqrt{2}$
that is independent of $m_X$ at leading order.
However, for scattering that is widely spread across a continuum of source
energies $E_\nu$, the width will be additionally correspondingly broadened.

We may test the prior conclusions against the numerical results in Fig.~\ref{fig:discLimits}.
For germanium, the minimal neutrino energy that can create a recorded recoil
(for head-on collisions, which have the lowest cross section) is around 0.5 or 2~MeV,
for threshold sensitivities \tmin of 10 or 100~eV, respectively.
Conversely, electron recoils can be sensitive to neutrinos as soft as a few keV.
In the nuclear recoil column, we observe a rather sharp transition knee,
centered at around 1 or 4~MeV for the baseline and future threshold
scenarios, which is stable across the different simplified models.
As expected, each is a bit larger (about a factor of two) than the
kinematic turn-on.  By eye the full width of the transition region is
approximately one order of magnitude in mass.
This suggests, consistently, that the continuum nature of the solar
and reactor sources is relevant here.
Looking at the electron scattering columns, we should expect a substantially
wider transition region, which starts at much lower masses, because the lower
reaches of the applicable neutrino energy spectra are not kinematically eclipsed.
We should observe, furthermore, that there is a less appreciable leftward shift
in the position of the knee for the second generation threshold sensitivity
since most relevant electron scattering events are already integrated at very
modest values of \tmin.  Based simply on the very wide range of electron
recoil energies $E_R$ which are sampled, spanning from threshold near
100~eV up to more than 10~MeV, we can expect a transition region around the
knee in $M_{Z'}$ that may cover more than two orders of magnitude,
from approximately 10~keV, up to a few MeV.  We additionally observe
more variation in the exact position and width of the knee
between the four BSM models in the electron scattering case.

In Fig.~\ref{fig:discLimits}, we plot the plateau sensitivity in the low mediator mass limit
as a function of the experimental recoil threshold.
As expected, going a lower threshold benefits the reach of
the experiments, especially in the low mediator mass region. This effect is more pronounced for nuclear
scattering since the electron scattering rate is flat with energy. The limiting factor of sensitivity to
electron scattering is the background (which is also flat), when it is removed the sensitivity increases
more than for nuclear scattering.
The nuclear recoil plots clearly exhibit a stairstep behavior in sensitivity for the Solar sources,
as the ${}^7{\rm Be}$ and pep line sources turn on at lower detection thresholds.
Comparing individual experiments for a given BSM scenario, the larger
flux of a reactor experiment allows for a greater reach when compared to an equivalent solar experiment.
However, this gap is smaller when considering electron scattering, since the pp rate is able to contribute
throughout the signal region, whereas for nuclei scattering the pp rate is below the detector threshold.
Additionally, comparing with a solar experiment performed in a background free environment, we find that
the reactor experiments will still have a greater reach in most BSM scenarios explored. These conclusions
are the same for the silicon experiments, the plots of which we have included in the appendix. In the
proposed configuration the silicon detectors are under-powered compared to germanium, owing to the
smaller detector mass and fewer neutrons per nuclei. The utility of the silicon component of the
experiment is in confirming a putative signal, not in adding statistical power to the overall
sensitivity. Having two detector materials would also provide complementary information in the event of
a discovery, for example, if there is isospin violation.

\begin{figure*}[ht]
\begin{tabular}{cc}
\includegraphics[height=5cm]{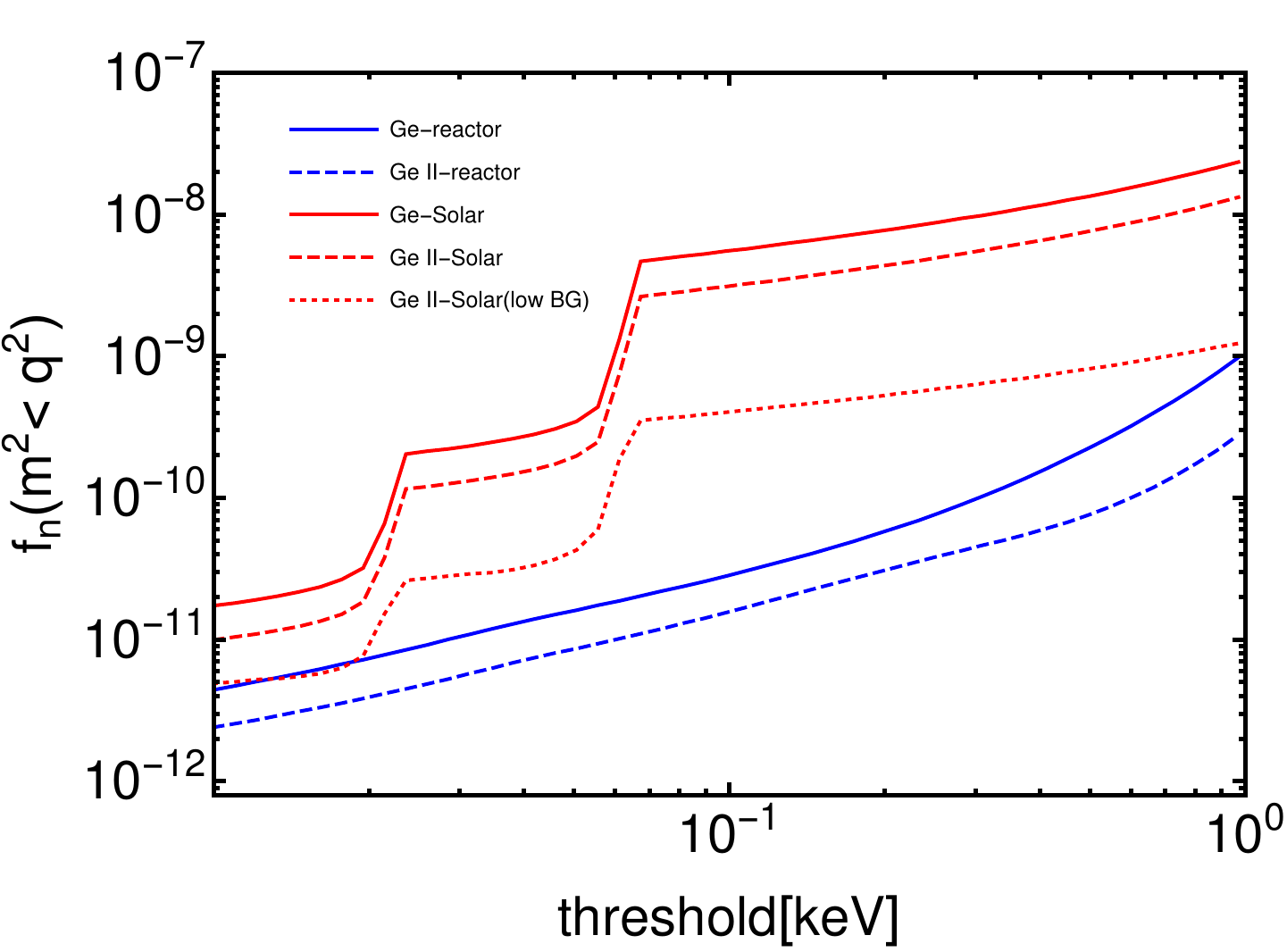}  &
\includegraphics[height=5cm]{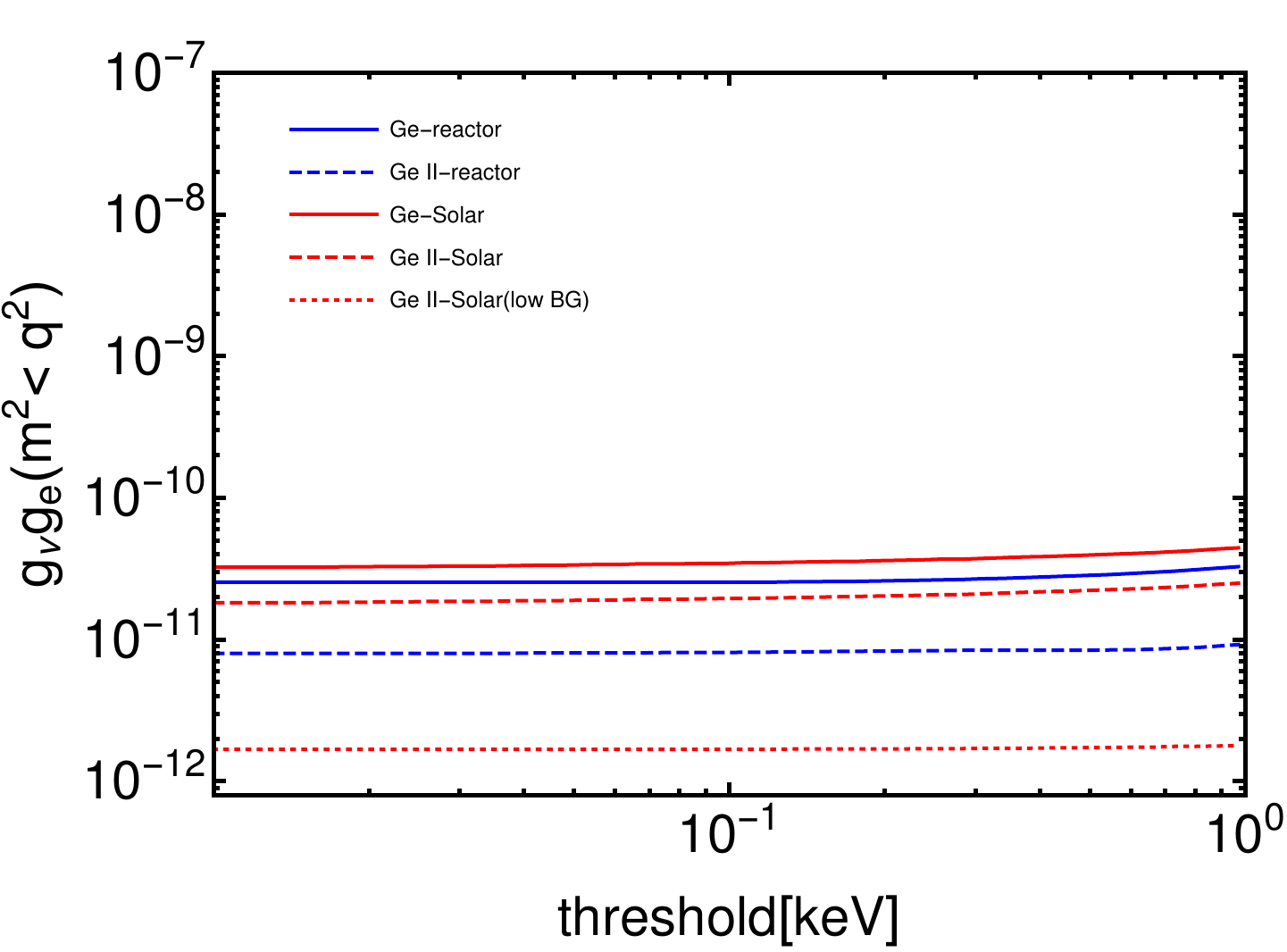} \\
 &
\includegraphics[height=5cm]{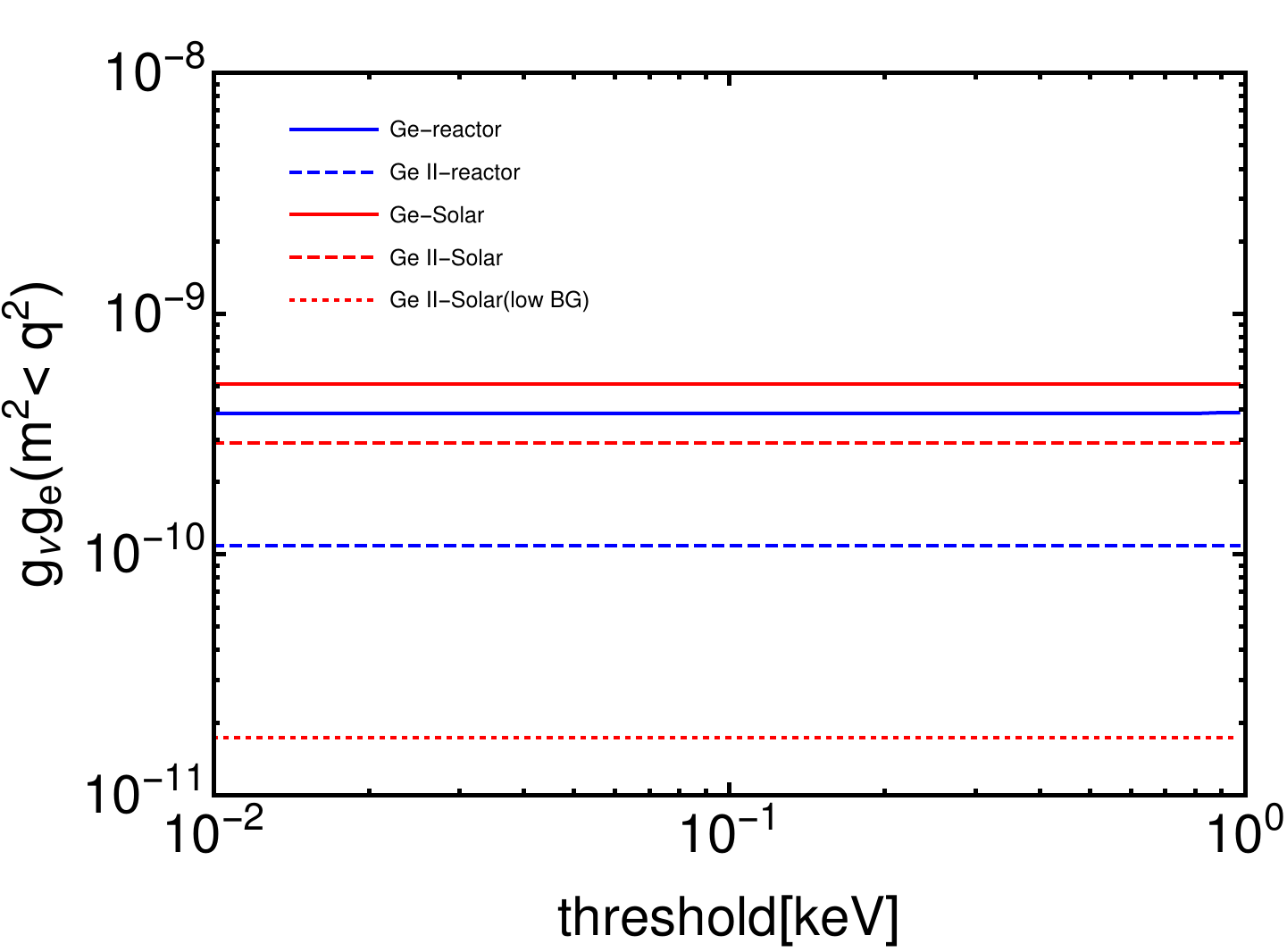} \\
\includegraphics[height=5cm]{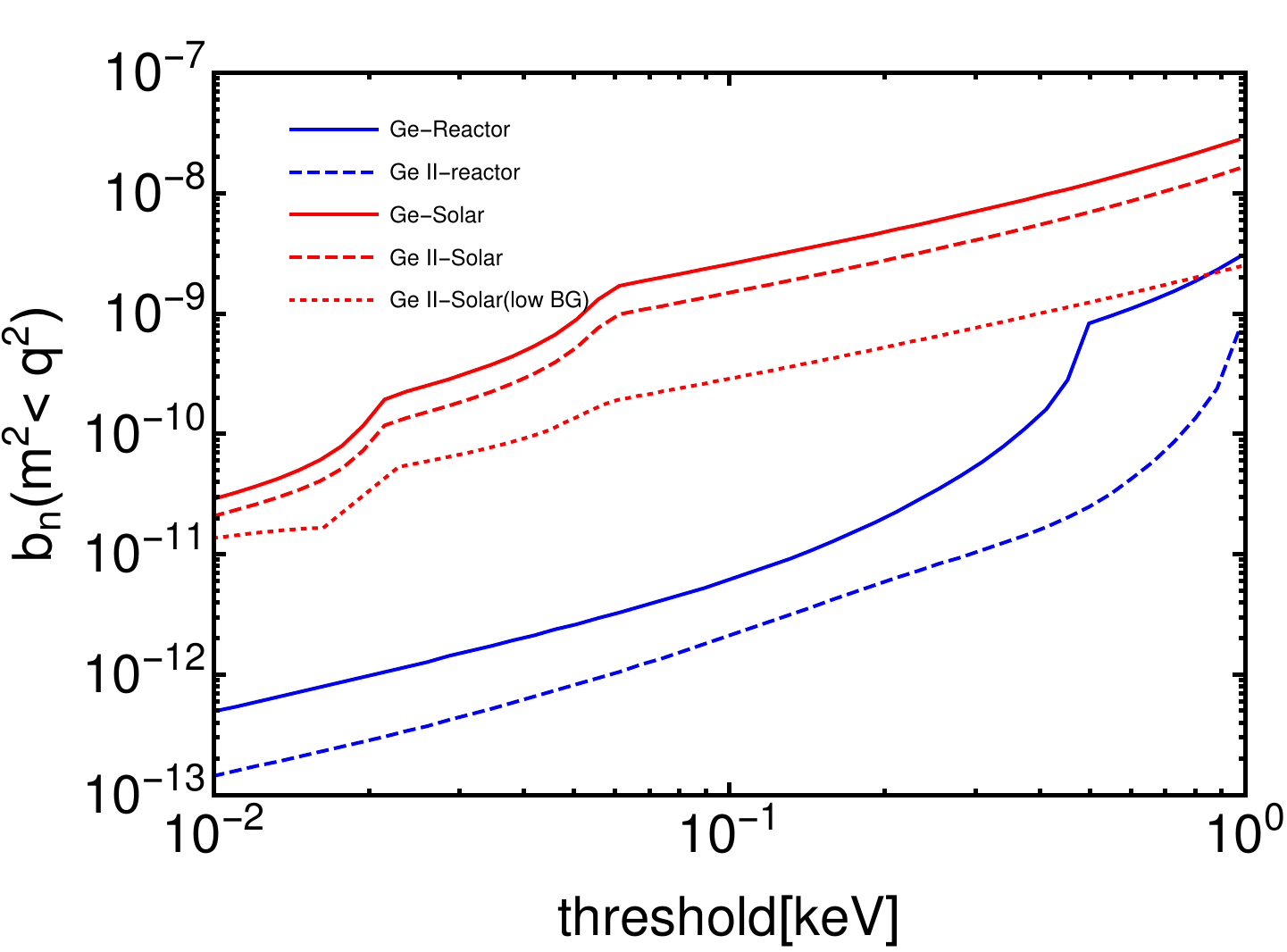}  &
\includegraphics[height=5cm]{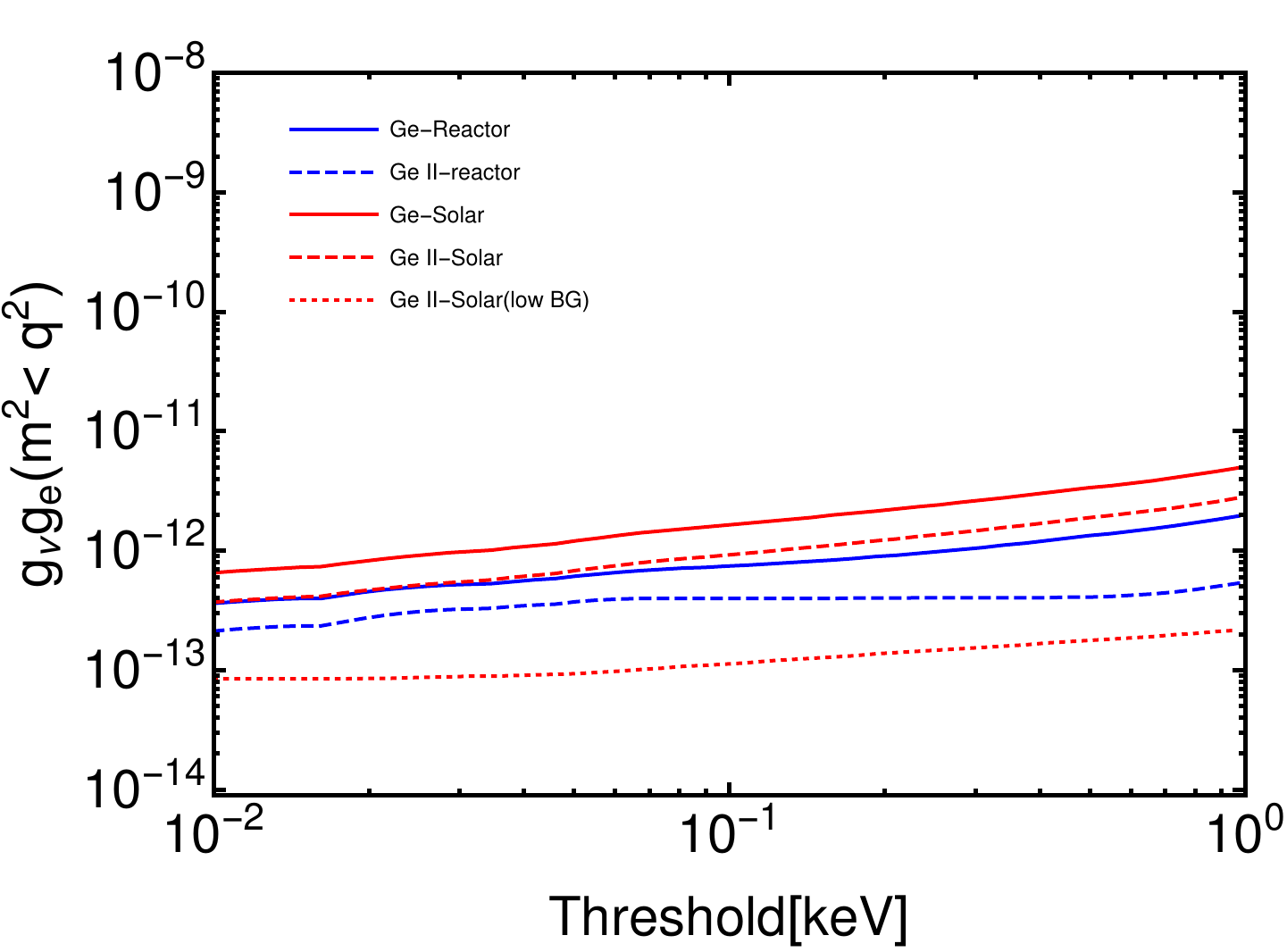} \\
\includegraphics[height=5cm]{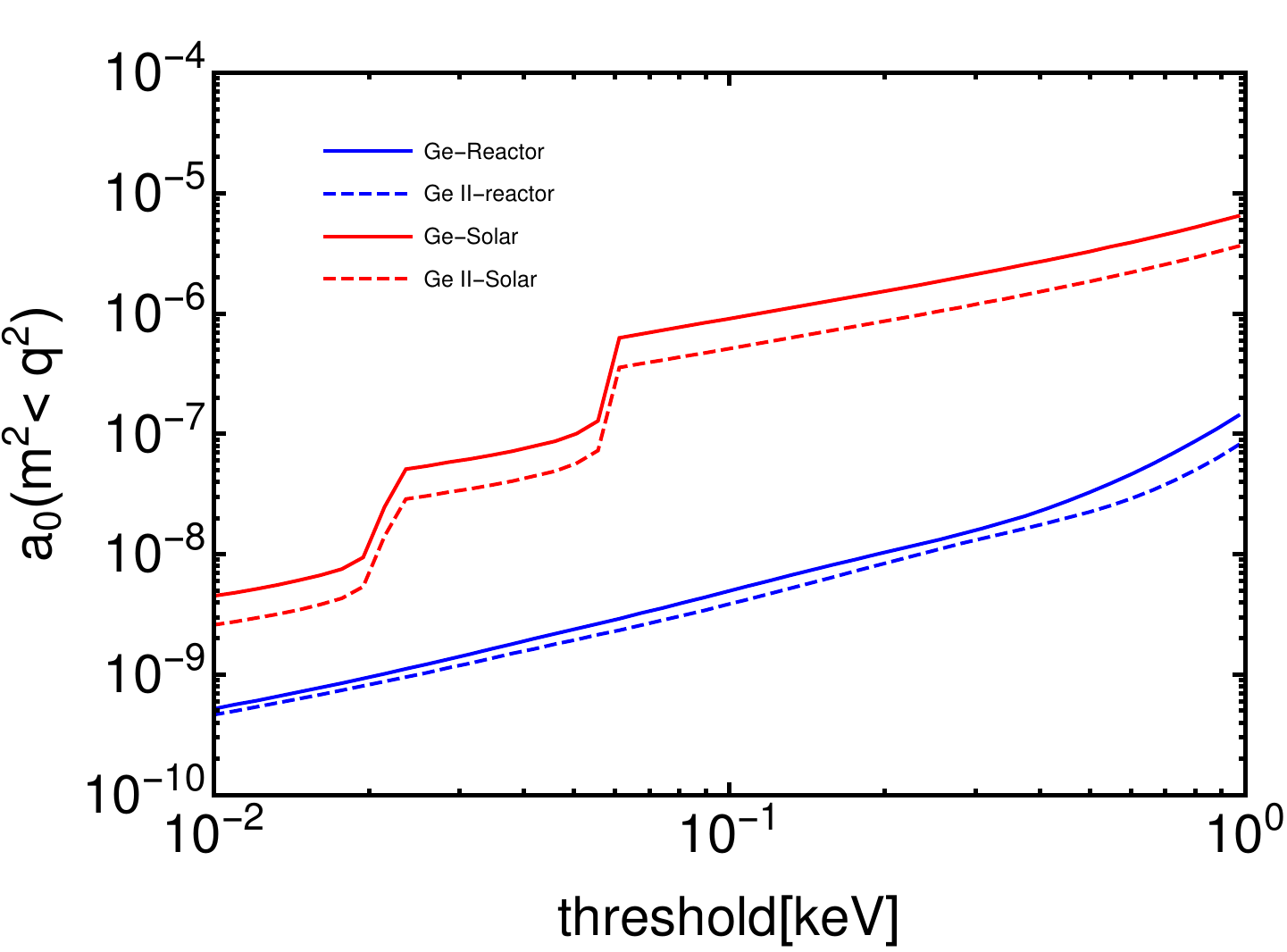}  &
\includegraphics[height=5cm]{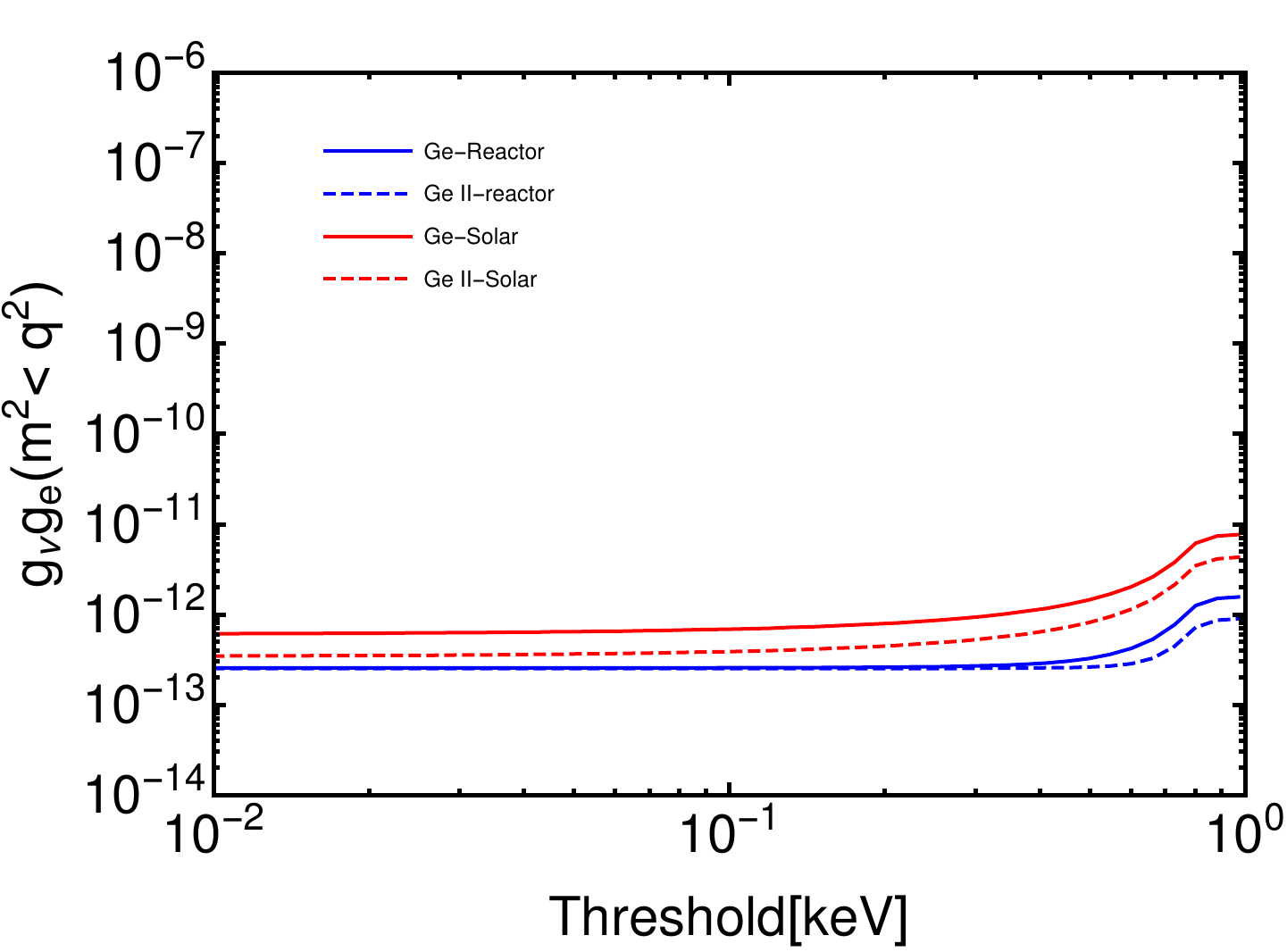} \\
\end{tabular}
\caption{Discovery limits vs. threshold for neutrino scattering off germanium nuclei (left) and electrons (right),
for the different BSM models (from top to bottom): scalar, pseudo-scalar, vector and axial-vector }
\label{fig:discLimitsTh}
\end{figure*}

\begin{figure*}[ht]
\includegraphics[height=9cm]{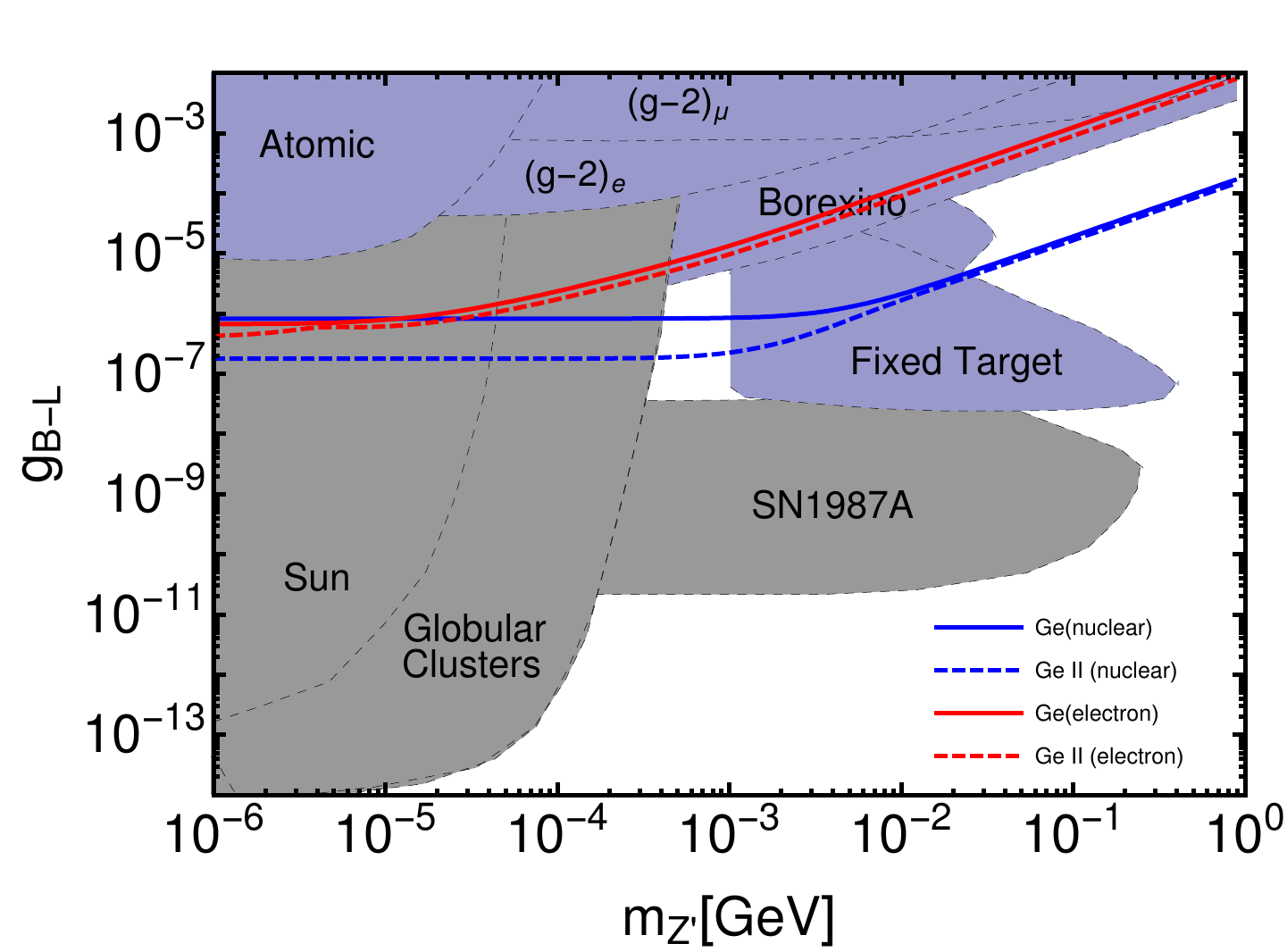}
\caption{Predicted exclusion curves in the coupling-mediator mass parameter space in a $U(1)_{\mathrm{B-L}}$ model
for electron (red) and nuclear (blue) recoils observed by germanium detectors (phase I and II are solid and dashed respectively) with a total
exposure of 10t-days, thresholds situated 2m from a $\bar{\nu}_e$ flux sourced by a 1MW reactor.
The shaded regions are exclusion curves from other experiments and observations, compilation from \cite{Harnik:2012ni}.}
\label{fig:BLexclusion}
\end{figure*}

To compare the reach of a reactor experiment with a more general set of constraints we examine a $U(1)_{\mathrm{B-L}}$
model with a light vector mediator of mass $m_{Z'}$ with coupling $g_{\mathrm{B-L}}$ described by the Lagrangian
\bea
\mathcal{L}_{\mathrm{B-L}} \supset && g_{\mathrm{B-L}}Z_{\mu}'\left(\frac{1}{3}\bar{q}\gamma^{\mu}q-\bar{\nu}\gamma^{\mu}\nu -\bar{e}\gamma^{\mu}e\right)
\eea
This model has also been explored for \cns sourced by solar neutrinos with G2 and future direct
detection experiments \cite{Harnik:2012ni,Cerdeno:2016sfi}. Fig.\ref{fig:BLexclusion} shows the
reach in the $g_{\mathrm{B-L}}-m_{Z'}$ space for the MINER scenario outlined above. The thick solid red curve gives
the reach due to nuclear scattering, and the thick solid blue curve is the reach for electron scattering.

A few aspects of this plot are worth detailing.  First, this demonstrates that a reactor experiment
employing low threshold detectors such as planned for MINER can provide world leading constraints in large
regions of parameter space.  Interestingly we find that the nuclear scattering constraints are greater
than those from electron scattering, which is a novel feature of the reactor experiment. Additionally,
the reach for nuclear scattering from the reactor is well below the reach for nuclear scattering from solar neutrinos. The blue and gray shaded regions are constraints from other experiments, whose details can be found for example in \cite{Harnik:2012ni,Cerdeno:2016sfi}.

\section{Scaling Rules\label{sct:scaling}}
In Sec.~\ref{sct:extensions}, we intuitively characterized the shape and scale
of contributions to the differential cross section for electron and
nuclear scattering from the exchange of a new light mediator.
In Secs.~\ref{sct:smconflimits} and \ref{sct:limits}, we outlined the statistical machinery by which
these profiles for new physics may be leveraged to set discovery limits,
and plotted limit curves in the coupling vs. mass plane for various
assumptions of the mediator spin, coupling type, background control,
and detector sensitivity.  In the present section, we deconstruct this
analysis approximately, but in closed form, in order to develop an intuitive
predictive framework for assessing experimental sensitivity.

In particular, we consider the asymptotic coupling sensitivity in the massless
mediator limit ($m_{Z'}^{2} \ll 2E_{R}m_X$), and attempt to establish rules for the relative scaling of
this limit with respect to variations such as the detector threshold \tmin,
the neutrino source (solar or reactor) and energy $E_\nu$,
the target mass (either an electron or a nucleus),
and the various BSM simplified models.
In particular, we will attempt to develop a framework whereby one
may approximately infer a family of related discovery limits
from knowledge of any one member.
In order to understand these relative sensitivities,
it is essential to inspect the way that the discovery limit is calculated,
in terms of the log-likelihood ratio $q_{0}$ defined in Eq.~(\ref{eq:loglike}).
After canceling the background contributions, and the nuisance parameter
terms (neglecting subleading model dependencies in their optimization), we observe
that any two models $A$ and $B$ will induce the same statistical deviation $\Delta q_0$
from the SM if $\log\mathcal{L}_{A}=\log\mathcal{L}_{B}$.  Since all discovery
limits will correspond to achieving a $\Delta q_0$ equal to certain Asimov number,
we can use this condition to solve for the desired scaling factors.

For simplicity, we will reduce the binned likelihood in Eq.~(\ref{eq:binnedlike}) to
a single integration.  Additionally, we notice that in the limit of large event counts,
the Poisson distribution reduces to the Gaussian with expected number $\nu$, observed
number $n$, and standard deviation $\sigma=\sqrt{\nu}$.
\begin{equation}
-2\log\mathcal{L}= \frac{\left(n-\nu\right)^{2}}{\nu}
\label{eq:logL}
\end{equation}
Note that ($n-\nu$) is the unified BSM event rate,
including any potential mixing terms with the SM.
We will have the same SM background rate $\nu$ for various models, so
the criterion for comparison of limits reduces in this approximation
simply to equality of (the absolute value of) the signal event deviation $\vert n-\nu \vert$,
or equivalently to equality of the magnitude of the BSM cross section
(allowing also for negative interference).

In order to extract simplified relationships on the new physics couplings $Q^{\prime}$,
we need to isolate the leading terms, focusing now specifically on the nuclear scattering example.
Given reasonable event rates and reliable background characterization,
experiments will be sensitive to small deviations from the SM rate.
The largest SM contributions are from the vector interaction,
and this will enhance the cross term for BSM vector mediators.
We define the coherent SM vector coupling to the nucleus as
$Q_v \equiv -2\times[Z g_v^p + (A-Z)g_v^n]$,
consistent with Ref.~\cite{Cerdeno:2016sfi}.
Correspondingly, we may neglect the square of the new physics term in this case.
By contrast, for axial vector interactions,
the mixing term with $Q_{v}$ is of order $\mathcal{O}(E_{R})$,
and the mixing term proportional to $Q_{a}$ is small because $Q_{a}\ll Q_{v}$.
Thus, we may neglect mixing in this case, and take the square of the new physics
amplitude to be leading.
For the scalar interactions, no cross term with the SM exists.
The associated cross-sections are calculable.
\begin{equation}
n-\nu
\,\, \appropto \,\,
\int_{\tmin} \hspace{-15pt} \mathrm{d}E_R
\,\,
\displaystyle \int_{(E_R+\sqrt{E_{R}^{2}+2E_{R}m_{N}})/2} \hspace{-82pt} \mathrm{d}E_{\nu} \hspace{65pt} f(E_{\nu})
\times
\begin{cases}
\frac{f_n^2 A^2 }{16\pi E_{\nu}^{2} E_{R}}
& \text{scalar}\\[6pt]
-\frac{b_n A G_{F} Q_{v}}{2\sqrt{2}\pi E_{R}}\left(1 - \frac{E_{R}}{\tmax} \right)
& \text{vector}\\[6pt]
\frac{a_0^2 {\langle S \rangle}^2}{8\pi m_N E_{R}^2}\left(1 + \frac{E_{R}}{\tmax} \right)
& \text{axial vector}
\end{cases}
\label{eq:n-v}
\end{equation}
where $f\left(E_{\nu}\right)$ is the anti-neutrino flux.
For a monochromatic beam, we may integrate in closed form, retaining leading terms.
\begin{equation}
\vert n-\nu \vert
\,\, \appropto \,\,
\begin{cases}
f_n^2 \times c_s \equiv \frac{A^2}{16\pi E_\nu^2} \ln\left(\frac{\tmax}{\tmin}\right)
& \text{scalar}\\[6pt]
b_n \times c_v \equiv \frac{A G_F Q_v}{2\sqrt{2}\pi} \left[\ln\left(\frac{\tmax}{\tmin}\right)-1\right]
& \text{vector}\\[6pt]
a_0^2 \times c_a \equiv \frac{{\langle S \rangle}^2 }{8 \pi m_N \tmin}
& \text{axial vector}
\end{cases}
\label{eq:n-vint}
\end{equation}
For a germanium target, we can take approximately $(A\to73)$, $(m_N \to 68~{\rm GeV})$, $(Q_v \to 38)$, and
$(\langle S\rangle \to 0.5)$, for ${}^{73}{\rm Ge}$ with isotopic fraction $({\rm IF}_{73} = 0.08)$.
Also, we have $G_F = 1.17\times 10^{-5}\,{\rm GeV}^{-2}$.  We will use $\tmin = 100$~eV,
corresponding to the first generation experiment, and select a representative neutrino
energy that is large enough to regularly trigger recoils above the threshold, say $E_\nu = 4$~MeV,
which implies $\tmax \simeq 470$~eV.  We may then estimate the relative limits
$(b_n/f_n^2 \simeq c_s/c_v \to 5\times10^9)$,
and $(b_n/a_0^2 \simeq c_a \times {\rm IF}_{73}/c_v \to 6\times 10^4)$, holding the event rate constant.
Taking the calculated reactor flux limit $(b_n = 10^{-12})$ for the vector coupling, we can predict
$(f_n \simeq 1\times 10^{-11})$ and $(a_0 \simeq 4\times 10^{-9})$ for the corresponding
scalar and axial vector limits, which agree very closely with the detailed calculations
illustrated in Fig.~\ref{fig:discLimits}.
Holding each of the Eq.~(\ref{eq:n-vint}) integrated rates constant, we can extract slopes for
variation of the coupling limits with respect to changes in the threshold \tmin.
\begin{equation}
\begin{cases}
\frac{d \log f_n}{d \log \tmin} \simeq \frac{1}{2 \ln\left(\tmax/\tmin\right) }
& \text{scalar}\\[6pt]
\frac{d \log b_n}{d \log \tmin} \simeq \frac{1}{\ln\left(\tmax/\tmin\right) - 1 }
& \text{vector}\\[6pt]
\frac{d \log a_0}{d \log \tmin} \simeq \frac{1}{2}
& \text{axial vector}
\end{cases}
\label{eq:n-slopes}
\end{equation}
A typical value of $\ln\left(\tmax/\tmin\right)$ is around 2.  This suggests that each of the plots
in the lefthand column of Fig.~\ref{fig:discLimitsTh} should have a log-slope of order approximately one,
steepest for the vector case, and trending steeper as one approaches the kinematic cutoff, which is
broadly consistent with what is observed in the detailed calculation.
The electron scattering plots are by contrast flat with respect to variation of the threshold.
This is not because the rate does not increase at low recoils, but rather because the event
rate in this region is typically dominated by nuclear recoils.  In order to carry out a corresponding
single bin treatment of electron scattering, one would need to select an integration region which
begins at a fixed scale around $\tmin \simeq 2$~keV,
where the \cns differential cross section tapers off and is eclipsed by the electron rate, as in Fig.\ref{fig:rates}. 
A possible exception to this conventional scenario may arise for a sufficiently light mediator,
if the electron rate is able to climb steadily in the keV region, to finally
compete with the nuclear rate in the exceptionally soft recoil region,
as described toward the end of Sec.~\ref{sct:extensions}. 

Next, we will briefly consider the relative strengths of limits achievable with solar and reactor
sources in the case of nuclear scattering.  Again, the strategy will be to hold the ratio $(n-\nu)^2/\nu$
constant.  We observe that both the BSM signal and SM background will scale linearly with the integrated
flux $\Phi$, such that the ratio to be held constant is likewise linear in $\Phi$.
A first generation experiment will be sensitive primarily to the ${}^8{\rm B}$ continuum, whose net
flux is diminished by a factor of $\mathcal{O}(10^5)$ relative to that of the reactor site under consideration
at the bare minimal baseline.  Additionally, however, there is sensitivity to the energy of the source,
with the event rate scaling as $E_\nu^2$.  The mean energy of this solar source is a factor of a few larger
than that of a terrestrial reactor, which offsets the rate advantage by approximately one order of magnitude,
to $\mathcal{O}(10^4)$, in the square.  The critical ratio $(n-\nu)^2/\nu$ is finally proportional to
$\Phi \times(f_n^4,b_n^2,a_0^4)$ in the scalar, vector, and axial vector models, after squaring the leading
signal contributions from Eq.~(\ref{eq:n-vint}).  This suggests that the solar limits should be weaker by
about a factor of $100$ in the vector case, or about $10$ in the scalar and axial cases.  This result
is complicated by the non-trivial response of each limit to background rates.
Indeed, if one compares the first generation solar and reactor limits
in Fig.~\ref{fig:discLimits}, the solar scattering disadvantage is more pronounced than predicted.
However, the lower-background second-generation limit ratios are broadly consistent with the simplified expectation. 
Similarly, as the recoil sensitivity threshold is reduced
sufficiently to access the ${}^7{\rm Be}$ and pep line sources
in Fig.~\ref{fig:discLimitsTh}, one sees discrete improvements in sensitivity consistent
with predictions around $5-6$ (scalar, axial vector) or $2-3$ (vector)
corresponding to escalation of the flux by a factor of around 30.

Finally, in the large mediator mass region, $2m_{N}E_{r}\ll m^{2}$,
we can estimate the order of $Q^{\prime}$ by comparing the BSM ``charge'' to the SM charge:
$\frac{1}{\sqrt{2}G_F}\frac{Q^{\prime}_{v}}{m^{2}_{z}}=\frac{1}{2}Q_{v}$,
so $Q^{\prime}=\frac{\sqrt{2}}{2}m^{2}G_{F}Q_{v}$. In order to distinguish BSM from
SM signal, one should need $Q^{\prime}$ larger than this value.

\section{Conclusion\label{sct:conclusions}}

Coherent elastic neutrino-nucleus scattering is a process with intriguing prospects for testing Standard Model physics and beyond.  Though its measurement has been an experimental target for many years, only recently has technological innovations made an observation realistically imminent. With various collaborations such as COHERENT, CONNIE, TEXONO, and MINER currently taking data or planning to take data in the very near future, and with the possibility of using direct dark matter detection to test CE$\nu$NS through detection of the solar neutrino flux, the time is ripe to determine the extent to which these various experimental avenues can explore new physics.

In the present work we have examined the effect of new, light mediating particles on both the CE$\nu$NS process as well as electron scattering by neutrinos, by employing a simplified model approach including new scalar, pseudoscalar, vector, and axial-vector mediators with sub-GeV masses. Such low mass mediators can create a substantial enhancement in the rate of CE$\nu$NS and $\nu-e$ scattering at low recoil energies, further motivating the continued push towards low threshold detector technology. 

Within this simplified model framework we have determined the projected reach of experiments using low-threshold germanium and silicon detectors at a distance of $\sim1-3$m from the core of a MW class nuclear reactor neutrino source (with the proposed MINER experimental configuration serving as a prototype), next generation Ge and Si direct dark matter detection experiments, as well as the currently running CsI detector deployed by the COHERENT group at the SNS at ORNL (the experimental parameters for these setups are given in Table \ref{tab:Detectors}). Following other recent studies in this area, we have adopted a $U(1)_{\mathrm{B-L}}$ BSM framework, and found that low-threshold Ge detectors at close proximity to a nuclear reactor have superior prospects for probing this model, as demonstrated in Fig.\ref{fig:BLexclusion}.  We  have also provided scaling rules that bolster our numerical results, which also provide an intuitive guide to the discovery potential of the CE$\nu$NS and $\nu-e$ processes in the coupling and mass parameter space when new light mediators are considered.

Experimental groups are poised on the cusp of a first-ever measurement of the CE$\nu$NS process. Theoretical efforts such as those detailed in the present work, combined with continued progress on the experimental front offer intriguing possibilities for near term results within this exciting new region of particle physics.

\section{Acknowledgements}
We thank Rupak Mahapatra for useful discussions. 
BD acknowledges support from DOE Grant DE-FG02-13ER42020.
LES acknowledges support from NSF grant PHY-1522717.
JWW acknowledges support from NSF grant PHY-1521105.
JBD and JWW thank the Kavli Institute for Theoretical Physics
for generous hospitality, hot coffee, and a most conducive view
during the early stages of this work.

\bibliography{PhysicsBibtex}

\end{document}